\documentclass{pasj00}

\begin{document}
\SetRunningHead{T. Nagayama et al.}{NH$_3$ emission in CMZ}
\Received{2008/12/04}

\title{NH$_3$ in the Galactic Center is formed in Cool Conditions}

\author{Takumi \textsc{Nagayama},\altaffilmark{1,6}
        Toshihiro \textsc{Omodaka},\altaffilmark{2} 
        Toshihiro \textsc{Handa},\altaffilmark{3} \\
        Hideyuki \textsc{Toujima},\altaffilmark{1,6} 
        Yoshiaki \textsc{Sofue},\altaffilmark{2}  
        Tsuyoshi \textsc{Sawada},\altaffilmark{4} \\
        Hideyuki \textsc{Kobayashi},\altaffilmark{5} 
        and Yasuhiro \textsc{Koyama}\altaffilmark{6}} 
\altaffiltext{1}{Graduate School of Science and Engineering, 
		 Kagoshima University, \\
                 1-21-35 K$\hat{o}$rimoto, Kagoshima 890-0065}
\altaffiltext{2}{Faculty of Science, Kagoshima University,
                 1-21-35 K$\hat{o}$rimoto, Kagoshima 890-0065}
\altaffiltext{3}{Institute of Astronomy, University of Tokyo, 
                 2-21-1 Osawa, Mitaka, Tokyo 181-0015}
\altaffiltext{4}{ALMA Project Office, 
                 National Astronomical Observatory of Japan, \\
                 2-21-1 Osawa, Mitaka, Tokyo 181-8588}
\altaffiltext{5}{Mizusawa VERA Observatory, 
	 	 National Astronomical Observatory of Japan, \\
		 2-21-1 Osawa, Mitaka, Tokyo 181-8588} 
\altaffiltext{6}{Kashima Space Research Center, National Institute of 
                 Information and Communications Technology,\\
                 893-1 Hirai, Kashima, Ibaraki 314-8501} 
\email{nagayama@astro.sci.kagoshima-u.ac.jp}

\KeyWords{Galaxy:center - Interstellar:molecules - Interstellar:ammonia} 

\maketitle

\begin{abstract}
It is an open question why the temperature of molecular gas
in the Galactic center region is higher than
that of dust. 
To address this problem, 
we made
simultaneous observations in the NH$_3$ $(J,K)$ = (1,1), (2,2), and (3,3) lines 
of the central molecular zone (CMZ) using the Kagoshima 6 m telescope. 
The ortho-to-para ratio of NH$_3$ molecules in the CMZ 
is 1.5--3.5 at most observed area.
This ratio is higher than the statistical equilibrium value,
and suggests that the formation temperature of NH$_3$ is 11--20 K.
This temperature is similar to the dust temperature
estimated from the submillimeter and infrared continuum.
This result suggests that the NH$_3$ molecules in the CMZ were produced 
on dust grains with the currently observed temperature (11--20 K),
and they were released into the gas phase by supernova shocks or collisions of 
dust particles.
The discrepancy between warm molecular gas and cold dust can be explained by
the transient heating of the interstellar media in the CMZ approximately
$10^5$ years ago when the NH$_3$ molecules were released from the dust.
\end{abstract}


\section{Introduction}

It is one of the biggest questions why the temperature of
molecular gas in the CMZ is much higher than that of dust.
Many molecular line observations have revealed that
the temperature of molecular gas is higher than
that of dust in the Galactic center region
(Nagayama et al. 2007, \cite{nag07} and references therein).
It is inconsistent with the standard molecular gas heating model
because there interstellar gas is mainly heated up by hot dust.
Surveys in NH$_3$ $(J,K)$ = (1,1) and (2,2) lines
have revealed that this inconsistency is not confined to small regions;
in fact, it extends over the entire molecular cloud complex in the Galactic center,
called the CMZ.
This implies that a gas heating process over a hundred parsec scale 
differs from that in the galactic disk.

To address this problem, 
the abundance ratio of ortho-NH$_3$ to para-NH$_3$, ortho-to-para ratio,
is a key.
An NH$_3$ molecule has two different structures.
One is called ortho-NH$_3$,
and the other para-NH$_3$
which are different in relative orientation of the
three hydrogen spins with respect to the nitrogen spin.
Ortho-NH$_3$ emits $K=3n$ $(n=0,~1,~2,~\cdots)$ lines
while para-NH$_3$ emits $K \neq 3n$ ($n=0,~1,~2,~\cdots$) lines.
Ortho-NH$_3$ and para-NH$_3$ are kept for a long time ($\sim 10^6$ yr) 
because no collisional or fast radiative process can transform between
these two configurations of NH$_3$.
The ortho-to-para ratio
is expected to be the statistical equilibrium value of 1.0
when NH$_3$ molecules are formed in the processes of 
gas phase reactions under high temperature.
The ortho-to-para ratio is expected to be larger than unity 
when NH$_3$ molecules adsorbed on cold dust grain surfaces are
released into the gas phase with excess desorption energy,
which is comparable to the energy difference between the ortho
and para states (\cite{ume99}).
Therefore, the ortho-to-para ratio indicates the
physical conditions in which NH$_3$ molecules are formed.
Moreover, NH$_3$ molecules formed on a dust grain surface
colder than 100 K cannot be released to interstellar gas
without the heating of dust (\cite{san93}).
NH$_3$ molecules with a low formation temperature
indicate that dust should be heated up by some process
such as the passage of a shock wave (\cite{ume99}).

In the CMZ,
the (3,3) emission is stronger than the (1,1) and (2,2)
emissions (\cite{mor83}).
This indicates that
ortho-NH$_3$ is overabundant as compared to para-NH$_3$.
However, the ortho-to-para ratio of NH$_3$ over the entire CMZ
has not yet been obtained.
We aim to clarify the formation mechanism of molecular gas in the CMZ
by simultaneous observations in NH$_3$ (1,1), (2,2), and (3,3) lines and
by estimating the ortho-to-para ratio.

We present large-scale maps of the CMZ in the NH$_3$ 
(1,1), (2,2), and (3,3) lines.
In section 2, we describe our observations in detail.
The observed data are presented in section 3.
In section 4, we discuss the physical conditions of the CMZ 
through measurements of the rotational temperature and 
the ortho-to-para ratio in NH$_3$ lines.  
In this paper, 
we assume that the distance to the Galactic center is 8.5 kpc
and we use the direction on the sky based on the galactic coordinates.  


\section{Observations}

\subsection{Data Obtained Using the Kagoshima 6 m Telescope}

We conducted a large-scale survey using the Kagoshima 6 m telescope of the
National Astronomical Observatory of Japan (NAOJ) from September 2005 to 
March 2007.
We made simultaneous observations in 
the NH$_3$ $(J,K)$ = (1,1), (2,2), and (3,3) lines 
at rest frequencies of 23.694495, 23.722633, and 23.870129 GHz,
respectively.
At 23 GHz, the telescope beamwidth was \timeform{9'.5} 
and the main beam efficiency ($\eta_{\mathrm{MB}}$) was 0.54.
We used a $K$-band HEMT amplifier whose system noise temperature was
200--300 K.
All spectra were obtained using a 8192-channel FX-type software-based spectrometer
developed by the Kagoshima University and 
the National institute of Information and Communications Technology (NiCT).
The bandwidth and frequency resolution of the spectrometer
are 256 MHz and 31.25 kHz, respectively.
At the NH$_3$ frequencies, these correspond to a velocity coverage
and velocity resolution of 3200 km s$^{-1}$ and 0.39 km s$^{-1}$, respectively.
We obtained 471 NH$_3$ (1,1), (2,2) and (3,3) spectra at
$\timeform{-1D.000} \leq l \leq \timeform{2D.000}$ and
$\timeform{-0D.375} \leq b \leq \timeform{+0D.375}$
with a spacing of \timeform{0D.125}.
The surveyed area corresponds to $450 \times 110$ pc. 
The NH$_3$ (1,1) and (2,2) data presented in this paper were
newly observed,
although we have already observed them, as shown in \cite{nag07}.
To obtain intensity ratios from (2,2) to (1,1) and (3,3) to (1,1) accurately,
we simultaneously observed three lines.
The data shown in \cite{nag07} are not included the analysis in this paper.

All data were obtained by position switching 
between the target positions and reference positions.
The reference positions are at the Galactic latitude 
$b < \timeform{-1D}$, 
where no NH$_3$ emission was detected.
We integrated at least 30 min at each point.
The relative pointing error was better than \timeform{1'}, which was 
verified by the observations of several H$_2$O (frequency; 22.235080 GHz) 
maser sources.

Data reduction was performed using the UltraSTAR
package developed by the radio astronomy group at the University of Tokyo
(\cite{nak07}).
To improve the signal-to-noise ratio, 
the obtained spectra are smoothed to a velocity resolution of 5 
or 10 km s$^{-1}$.
The rms noise level after smoothing to 5 km s$^{-1}$ is typically 0.030 K
in unit of the main beam brightness temperature defined by  
$T_{\mathrm{MB}} \equiv T_{\mathrm{A}}^{*} / \eta_{\mathrm{MB}}$, 
where $T_{\mathrm{A}}^{*}$ is the antenna temperature 
calibrated by the chopper wheel method (\cite{kut81}).
This sensitivity is better than that in the data shown in \cite{nag07}
by a factor of 2.7.
In this paper, the intensities are presented in the main beam 
temperature.

\subsection{Data Obtained Using the Kashima 34 m Telescope}

In order to conduct an investigation 
of the ``\timeform{0D.9} wing feature'',
which is a high-velocity wing at $l \sim \timeform{0D.9}$
at higher resolution (see section 4.5.5), 
we observed the area at
$\timeform{0D.700} \leq l \leq \timeform{0D.940}$ and
$\timeform{-0D.180} \leq b \leq \timeform{+0D.020}$
with a spacing of \timeform{0D.040}.
This observation was made using the \timeform{1'.6} beam of 
the Kashima 34 m telescope of NiCT
from May to November in 2007.
The spectra in the NH$_3$ (1,1), (2,2), and (3,3) lines 
were obtained by the same method as that used 
in the Kagoshima 6 m telescope survey.
We integrated at least 5 min at each point.
The rms noise level after 5 km s$^{-1}$ smoothing is typically 0.064 K
in $T_{\rm MB}$.


\section{Results}

\subsection{Profiles}

\begin{figure*}[h]
  \begin{center}
    \FigureFile(160mm,80mm){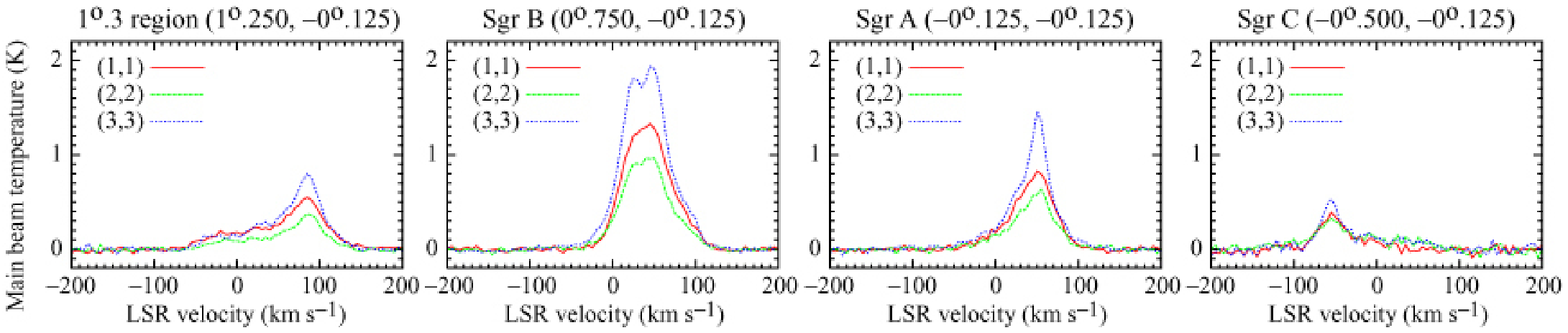}
  \end{center}
  \caption{The NH$_3$ (1,1) (red), (2,2) (green), and (3,3) (blue)
           spectra toward four positions of the observed area,
           Sgr C $(l,~b) = (\timeform{-0D.500},~\timeform{-0D.125})$,
           Sgr A $(\timeform{0D.125},~\timeform{-0D.125})$,
           Sgr B $(\timeform{0D.750},~\timeform{-0D.125})$, and
           \timeform{1D.3} region $(\timeform{1D.250},~\timeform{-0D.125})$.
           Spectra are shown with the veolcity resolution of 5 km s$^{-1}$
           and on a $T_{\mathrm{MB}}$ scale.}
  \label{fig:1}
\end{figure*}

Direct information about the observed lines can be seen in each line profile.
Significant NH$_3$ (1,1), (2,2), and (3,3) emissions for which
the signal-to-noise ratio exceeds 3 were detected for 101, 84, and 97
positions out of 157 observed positions, respectively.
The observed profiles in the (1,1), (2,2), and (3,3) lines 
toward four prominent positions
[Sgr C $(l,~b) = (\timeform{-0D.500},~\timeform{-0D.125})$,
Sgr A $(\timeform{0D.125},~\timeform{-0D.125})$,
Sgr B $(\timeform{0D.750},~\timeform{-0D.125})$, and
\timeform{1D.3} region $(\timeform{1D.250},~\timeform{-0D.125})$]
are shown in Figure \ref{fig:1}.
The (2,2) line is the weakest, and
the (3,3) line is the strongest at most positions,
especially near the velocity of peak intensity.

The line shapes of the three transitions are similar.
Although an NH$_3$ line profile has five quadruple hyperfine lines,
we cannot separate these components due to the large internal motion
of the gas in an observed beam.
The line shapes reflect the violent motion of the gas .

\subsection{Intensity Distributions over the Entire CMZ}

\begin{figure*}[h]
  \begin{center}
    \FigureFile(160mm,160mm){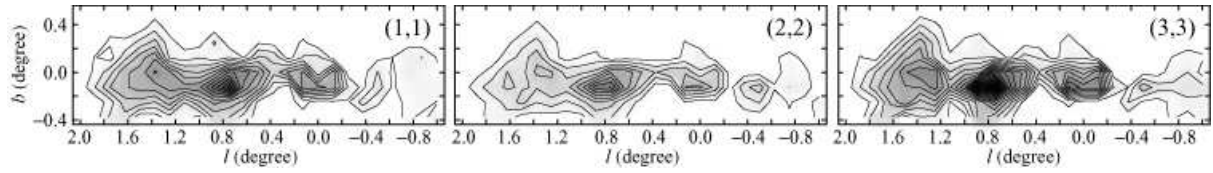}
  \end{center}
  \caption{Integrated intensity maps of NH$_3$ (1,1) (left), (2,2) (middle), 
           and (3,3) (right). 
           The lowest contour and the contour interval are 
           2,2 and 4.4 K km s$^{-1}$ in $\int T_{\mathrm{MB}} dv$, respectively.
           The velocity range of the integration is 
           $-200 \leq v_{\mathrm{LSR}} \leq 200$ km s$^{-1}$.}
  \label{fig:2}
\end{figure*}

To figure out the NH$_3$ gas distribution on the sky 
over the entire CMZ,
the velocity integrated intensity distribution is useful
for comparison with the maps observed in radio and infrared continuum emissions.
Figure \ref{fig:2} shows the $l$-$b$ maps of intensities, velocity-integrated
in the range of $v_{\rm LSR} = -200$ to 200 km s$^{-1}$,
covering the NH$_3$ emission.

The observed area is extended to $l > \timeform{1D.6}$ from
the previous survey (\cite{nag07}) along the galactic plane.
The distributions in the (1,1) and (2,2) lines are consistent with
those shown in \cite{nag07};
in addition, fainter peaks and diffuse emission were detected because of
the improved sensitivity.
The total integrated intensities in the (1,1), (2,2), and (3,3) lines
of the entire observed area are
1693, 1159, and 2120 K km s$^{-1}$, respectively.

The (3,3) line is emitted by the ortho-NH$_3$.
The (1,1) and (2,2) lines are emitted by the para-NH$_3$.
Ortho and para-NH$_3$ cannot be transformed by 
collisional or fast radiative process in the interstellar gas.
This implies that the ortho- and para-NH$_3$ molecules 
can be dealt with as different molecular species.
Therefore, the similarity of the intensity distribution in the (3,3) line 
with those in the (1,1) and (2,2) lines
should contain some astrophysical information.

\begin{figure*}[h]
  \begin{center}
    \FigureFile(160mm,100mm){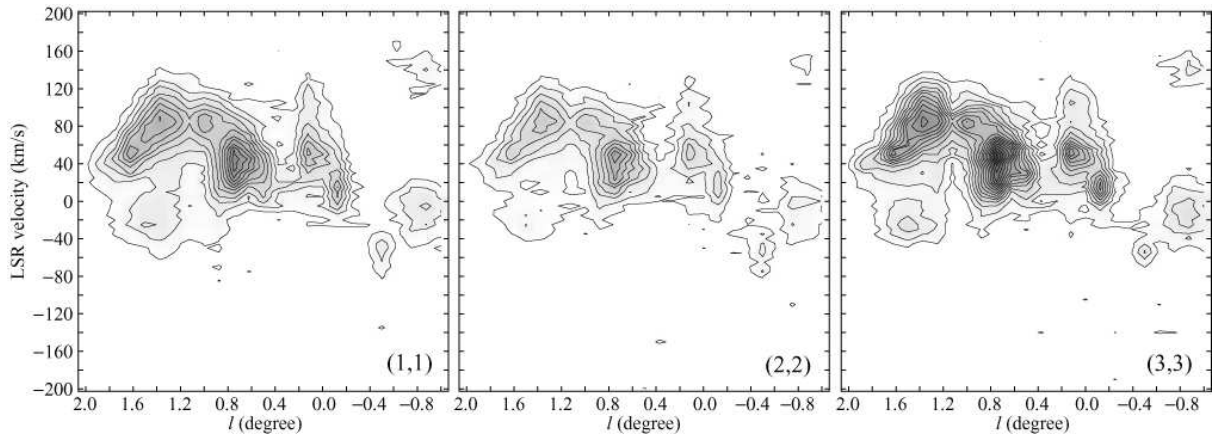}
  \end{center}
  \caption{Longitude-velocity ($l$-$v$) diagrams of NH$_3$ (1,1) (left), (2,2) (middle),
           and (3,3) (right) emission integrated over the entire observed latitude.
           The lowest contour and the contour interval are 0.25 and 0.31 K, respectively.}
  \label{fig:3}
\end{figure*}

The $l$-$b$ maps in the NH$_3$ lines appear similar to 
other molecular lines such as those of CO (\cite{saw01}) and CS (\cite{tsu99}).
This similarity is not only in the sky distribution but in the $l$-$v$ domain.
Figure \ref{fig:3} shows 
$l$-$v$ diagrams of three transitions integrated over the observed latitudes
and Figure \ref{fig:4} shows the $l$-$v$ diagrams at a fixed latitude.
Even for $v_{\rm LSR} = 0$ km s$^{-1}$, 
the NH$_3$ lines are not affected by
foreground absorption; in contrast, the effect is significant in the CO line.
Due to the difference of the critical densities, 
the CO line traces lower-density gas whereas the NH$_3$ line traces
only higher-density gas, $n$(H$_2$) $\sim 10^4$ cm$^{-3}$.
The NH$_3$ and CO lines trace the different density gas microscopically.
Therefore, morphological similarity suggests that the flux ratio of core
to envelope in an observed beam is almost uniform over the CMZ.
Moreover, the fairly large optical depths in both lines ($\tau \lesssim$ 2--4)
may reduce the morphological difference, because both lines mainly trace the outer envelope
of the cloud complex.


\subsection{Observational Properties of Individual Clouds}

To investigate the gas properties at various locations in the CMZ,
we divide the intensity distribution into several small regions and
define a {\it cloud} as 
a single peak in the $l$-$v$ and $b$-$v$ distributions,
with the intensities
stronger than the 3 $\sigma$ level 
($> 0.1$ K in $T_{\rm MB}$) in all three lines.
Based on this definition, 
we identified 13 clouds. 
The total integrated intensity of these 13 clouds is 
3447 K km s$^{-1}$,
which corresponds to
69\% of the integrated intensity of the entire observed area.
Therefore, the majority of the NH$_3$ gas in the CMZ 
is traced by these clouds.
The peak positions, 
apparent sizes, and line widths of these clouds are listed in Table \ref{tab:1}.
We defined their sizes and widths 
($\Delta l$, $\Delta b$, and $\Delta v$)
as apparent full widths at half-maximum (FWHM) of the main-beam temperature
without beamsize deconvolution.
The integrated intensities of the clouds are shown in Table \ref{tab:2}.

Four major clouds of the CMZ are
seen in the NH$_3$ maps
(from west to east:
the Sgr A 20 km s$^{-1}$ cloud at $l \simeq \timeform{-0D.1}$,
the Sgr A 40 km s$^{-1}$ cloud at $l \simeq \timeform{0D.1}$,
the Sgr B cloud at $l \simeq \timeform{0D.7}$, and
the \timeform{1D.3} region cloud at $l \simeq \timeform{1D.3}$).
These clouds 
were also detected in the previous observation (\cite{nag07}),
and their locations and extents are found to be the same with 
those of the previous ones.
In our new map, we clearly detected the NH$_3$ clouds 
associated with the Sgr C H\emissiontype{II} region at 
$(l,~b,~v_{\rm LSR}) \simeq (\timeform{-0D.5},~\timeform{-0D.1},~-50~\rm{km~s}^{-1})$ and
a part of the expanding molecular ring (EMR) at
$(l,~b,~v_{\rm LSR}) \simeq (\timeform{-0D.9},~\timeform{-0D.1},~140~\rm{km~s}^{-1})$.
Both had been marginally detected in the previous observation.


\subsection{Intensity Ratios}

The intensity ratio of the (2,2) line to the (1,1) line, $R_{(2,2)/(1,1)}$,
is determined by the optical depth, $\tau$,
and the kinetic temperature, $T_{\rm k}$, of the gas.
With a given optical depth,
the rotational temperature of the transition, 
$T_{\rm rot}$, can be derived from the $R_{(2,2)/(1,1)}$
and it gives the gas kinetic temperature.
Therefore, the changing line ratios indicate the differences of the rotational temperature.

Figure \ref{fig:5}(a) shows the histograms of
ratio for the integrated intensity over the entire observed area 
with 3 $\sigma$ detection in both the (1,1) and (2,2) lines 
after smoothing to 10 km s$^{-1}$.
The value of $R_{(2,2)/(1,1)}$ ranges between 0.4 and 1.5.
The mean value and standard deviation of $R_{(2,2)/(1,1)}$
are derived to be 0.70 and 0.15, respectively.
The NH$_3$ integrated intensity with
$R_{(2,2)/(1,1)} = 0.4$--0.8 and
$R_{(2,2)/(1,1)} > 0.8$ are
82\% and 18\% of the total NH$_3$ integrated intensity, respectively.
These values are consistent with the results described in \cite{nag07}.

The intensity ratio of the (3,3) line to
the (1,1) line, $R_{(3,3)/(1,1)}$,
is determined by the relative abundance of the ortho-to-para NH$_3$, $R_{\rm o/p}$,
and the kinetic temperature.
In the case that ortho-NH$_3$ and para-NH$_3$ have the same kinetic temperature,
$R_{(3,3)/(1,1)}$ increases with $R_{\rm o/p}$.
In the case that the ortho-to-para ratio is uniform, 
$R_{(3,3)/(1,1)}$ increases with the kinetic temperature.
We can estimate the ortho-to-para ratio using the kinetic temperature obtained by the $R_{(2,2)/(1,1)}$.
The ortho-to-para ratio depends on the temperature
when NH$_3$ molecules are formed.
The ortho-to-para ratio increases in the case that
NH$_3$ molecules are produced in low temperature conditions
(e.g., more than 10 at 5 K) (\cite{tak02}).
$R_{(3,3)/(1,1)}$ is the direct observed value 
that is related to the ortho-to-para ratio,
and therefore, it should be studied in detail. 

The (3,3) emission is stronger than the (1,1) emission
at most observed positions.
The NH$_3$ integrated intensity for $R_{(3,3)/(1,1)} > 1$ is
90\% of the total NH$_3$ intensity.
The histogram of $R_{(3,3)/(1,1)}$ is shown in Figure \ref{fig:5}(b).
The mean value and standard deviation
of $R_{(3,3)/(1,1)}$ are derived to be 1.28 and 0.29, respectively.
$R_{(3,3)/(1,1)}$ of the CMZ is higher than 
that of typical molecular clouds in the galactic disk.
In galactic disk clouds,
$R_{(3,3)/(1,1)}$ as high as that observed in the CMZ is found only 
in active and massive star-forming regions 
(see section 4.2).

To investigate the spatial distributions 
of $R_{(2,2)/(1,1)}$ and $R_{(3,3)/(1,1)}$,
we show the $l$-$v$ diagrams of $R_{(2,2)/(1,1)}$ and $R_{(3,3)/(1,1)}$ 
superimposed on the (1,1) emission integrated over the entire observed latitude
in Figure \ref{fig:6}.
The values of $R_{(2,2)/(1,1)}$ and $R_{(3,3)/(1,1)}$ 
vary with the location.
The high-ratio region is common in both the ratios.
Although the kinematical timescale significantly depends on
the galactocentric distance,
neither ratio exhibits any systematic structure dependent on
the distance from the Galactic center. 
Overall, the distributions of $R_{(2,2)/(1,1)}$ and $R_{(3,3)/(1,1)}$ 
are similar.
However, some regions
such as the Sgr C cloud near $l \simeq \timeform{-0D.5}$
exhibit a much lower $R_{(3,3)/(1,1)}$. 


\section{Discussion}

\subsection{Physical Properties of Molecular Gas in CMZ}

$T_{\rm k}$ derived from the $R_{(2,2)/(1,1)}$
is thought to be the current temperature of the gas.
The temperature derived from the ortho-to-para ratio 
corresponds to the temperature when the NH$_3$ molecules were formed
(\cite{ume99}).
These two temperatures can be obtained from the NH$_3$ observations.
In the case that the entire interstellar matter is under thermal equilibrium 
and in a steady state,
the two temperatures are the same.

We derived the $T_{\rm k}$ of the gas 
under the 
cases of optically thin ($\tau \ll 1$) and thick ($\tau \sim 3$)
using the same method as that described in \cite{nag07}.
The gas is warmer than the dust ($T_{\rm dust} = 15$--22 K; \cite{lis01})
at most observed positions, which is consistent with the conclusion shown in \cite{nag07}.
The $R_{\rm (2,2)/(1,1)} = 0.4$--0.8 and 
$R_{\rm (2,2)/(1,1)} > 0.8$ correspond to
$T_{\rm k} \simeq 20$--80 K and $T_{\rm k} > 80$ K, respectively.
The gases corresponding to $T_{\rm k}$ values of 20--80 K and $> 80$ K contain
82\% and 18\% of the total NH$_3$ integrated intensity 
in the entire observed area, respectively.
The higher ratio ($R_{(2,2)/(1,1)} \geq 0.7$) components are located
at $l \simeq \timeform{-0D.5}$, \timeform{0D.0}, \timeform{0D.9}, and \timeform{1D.3}
(Figure \ref{fig:6}(a)).
These locations coincide with
those of high CO $J =$ 3--2/$J =$ 1--0 ratios
($R_{\rm CO 3-2/1-0} \geq 1.5$; \cite{oka07}).
This supports the notion that
the components with higher $R_{(2,2)/(1,1)}$ values trace the high-temperature gas.
$R_{(2,2)/(1,1)} \geq 0.7$ corresponds to 
$T_{\rm k} \geq 60$ K in the optically thin case and 
$\geq 40$ K in the optically thick case ($\tau \sim 3$).
This is consistent with $R_{\rm CO 3-2/1-0} \geq 1.5$ corresponding to
$T_{\rm k} \geq 48$ K, 
which is derived from a large velocity gradient (LVG) calculation (\cite{oka07}).

The high $R_{(3,3)/(1,1)}$ value in the CMZ indicates
that the ortho-NH$_3$ is overabundant as compared to para-NH$_3$.
To obtain the ortho-to-para ratio,
we estimated the column densities using
the rotation diagram method under the optically thin assumption.
We derived the column densities of each level from the integrated intensities.
The procedure is based on \citet{tak02}.
Although the optical depth of the NH$_3$ lines in the CMZ
is moderate ($\tau \leq$ 2--4; \cite{hut93}; \cite{nag07}).
and the ortho-to-para ratio changes with the optical depth,
the value obtained under the optically thin assumption 
can provide the lower limit of the ortho-to-para ratio
($R_{\rm o/p} = 2$ at $\tau \ll 1$ corresponds to $R_{\rm o/p} = 6$ at $\tau = 3$).

Because of the absence of emission lines from molecules at the (0,0) level,
we cannot obtain the column density of the (0,0) level directly.
However, we can derive the column density of the NH$_3$ at the (0,0) level
under the assumption that the ortho- and para-NH$_3$ molecules are
thermalized at the same rotational temperature derived from the $R_{(2,2)/(1,1)}$.
This assumption is a better approximation
than using the rotational temperature between (3,3) and (6,6),
because 
the (1,1) and (2,2) levels are between the (0,0) and (3,3) levels.
The energies of the (3,3) and (6,6) levels are 
124.5 K and 412.4 K above the ground state, respectively.
The large energy difference between the (3,3) and (6,6) levels
suggests that the (3,3) and (6,6) emissions originate from different components.

We calculated the column densities of ortho-NH$_3$ and para-NH$_3$ by
the formulas $N{\rm (ortho)} = N(0,0) + N(3,3)$ and
$N{\rm (para)} = N(1,1) + N(2,2)$, respectively.
The column densities of the levels higher than (4,4)
are less than 2\% of those of the (0,0) and (1,1) levels
when the rotational temperature is 36 K,
which is estimated from the mean $R_{(2,2)/(1,1)} = 0.70$.
We considered only the metastable states ($J = K$),
because transitions from non-metastable states ($J \neq K$)
to metasable states are generally quite rapid.

Figure \ref{fig:7} shows histograms of
the ortho-to-para ratio for the integrated intensity 
over the entire observed area
with 10 $\sigma$ detection in the (1,1), (2,2), and (3,3) lines
after smoothing to 10 km s$^{-1}$.
The mean value and the standard deviation of $R_{\rm o/p}$
are derived to be 2.0 and 0.6, respectively.
To investigate the spatial distributions
of the ortho-to-para ratio,
we have plotted the ortho-to-para ratio
on the $l$-$v$ plane, as shown in Figure \ref{fig:8}.
We found that 
the ortho-to-para ratio varies with the location, 
and it does not exhibit systematic structures dependent on 
the distance from the Galactic center nor the core/envelope of clouds.

The distributions of $R_{(2,2)/(1,1)}$ and $R_{(3,3)/(1,1)}$
appear similar on the $l$-$v$ plane (Figure \ref{fig:6}).
To see this correlation quantitatively, 
we made a correlation plot of $R_{(2,2)/(1,1)}$ 
and $R_{(3,3)/(1,1)}$, as shown in Figure \ref{fig:9}.
$R_{(3,3)/(1,1)}$ depends on both $R_{\rm o/p}$ and $T_{\rm rot}$,
although $R_{(2,2)/(1,1)}$ depends on only $T_{\rm rot}$.
Therefore, we can draw curves of constant $R_{\rm o/p}$ values
on the $R_{(2,2)/(1,1)}$-$R_{(3,3)/(1,1)}$ domain (Figure \ref{fig:9}).
Figure \ref{fig:9} shows the curves with $R_{\rm o/p} =$ 1, 2, 4, and 6.
Figure \ref{fig:9} shows 
also a plot of the observed ratios of the clouds in the galactic disk
obtained from other observations.
Most galactic disk clouds such as M17, Cep A, and NGC 7538,
show $R_{\rm o/p} \simeq$ 1,
although intense and massive starforming cores
such as W49 and W51 are located
along the curve with $R_{\rm o/p} \simeq 4$.
The CMZ is close to the latter and 
is different from the former.
This suggests that the entire CMZ should be 
under physical conditions
as similar to an intense starforming core.

It has been reported that the ortho-to-para ratios of 
nearby external galaxies are as high as that of the CMZ.
The ortho-to-para ratios of Maffei 2, Arp 220, and NGC 253 are
2.6, 1.7, and 6.2, respectively (\cite{tak00}; 2002; 2005).
We calculated the orhto-to-para ratios of IC 342 and M 82
using the data of \citet{mau03} and \citet{wei01}, respectively.
These have been plotted in Figure \ref{fig:9}.
From these data, IC 342 exhibits
lower $R_{\rm o/p}$ as compared to that of other galaxies. 
However, $R_{\rm o/p}$ of IC 342 is calculated to be $1.8 \pm 0.5$
from the data obtained using the NRO 45 m telescope
(S. Takano, 2008 private communication).
Therefore, IC 342 may actually have a high $R_{\rm o/p}$ value.

M82 also shows low $R_{\rm o/p}$. 
\citet{use07} suggests that
the evolutionary stage of the starburst in M 82
is different from those of NGC 253 and IC 342.
This difference may relate to the low ortho-to-para ratio in M 82.

The ortho-to-para ratio is $\gtrsim 2$
for the CMZ and most of external galaxies.
We conclude that the ortho-to-para ratio of the CMZ 
and the central region of the galaxies is higher than that of galactic disk.


\subsection{Origin of NH$_3$}

The ortho-to-para ratio is related to the temperature
when NH$_3$ molecules are formed.
The ortho-to-para ratio approaches unity
when NH$_3$ molecules are produced under high temperature ($\geq 40$ K) conditions.
This ratio increases
when NH$_3$ molecules are produced 
and equilibrated at low temperature (e.g., more than 10 at 5 K).
This is because most of molecules then reside in the lowest state,
or the (0,0) ortho level.
The relationship between the ortho-to-para ratio and the
formation temperature is shown in Figure 3 of \citet{tak02}.
Our observation shows the ortho-to-para ratio
is 1.5--3.5 for 95\% of the total 
integrated intensity of the entire CMZ (Figure \ref{fig:7}).
This ratio corresponds to a formation temperature of 11--20 K.
This is much lower than 
the gas kinetic temperature derived from $R_{(2,2)/(1,1)}$.

The low formation temperature of NH$_3$ suggests that
most of the NH$_3$ in the CMZ was not produced by a gas-phase reaction.
If NH$_3$ were produced with a formation temperature of $\geq 20$ K,
that is, the current gas temperature,
the ortho-to-para ratio should be $< 1.5$.
Only 5\% of the total integrated intensity shows $R_{\rm o/p} < 1.5$.

It should be noted that
the formation temperature of NH$_3$ is similar to the dust temperature in the CMZ, 
which is estimated to be 
$\simeq$ 15--22 K (\cite{lis01}).
This suggests that the observed 
NH$_3$ molecules were formed on the surface of cold dust grains.

At the observed formation temperature, i.e., 11--20 K,
NH$_3$ molecules cannot be released from the dust surface into a gas phase.
Therefore, the dust should be heated up to 100 K
to release the NH$_3$ from the dust to the gas phase,
because the ice at 100 K can release NH$_3$ within a few years (\cite{san93}).
The transition timescale of ortho-NH$_3$ and para-NH$_3$ is 
$\sim 10^6$ yr in the interstellar gas (\cite{che69}).
Since the observed ortho-to-para ratio is larger than the statistical equilibrium value,
NH$_3$ would be released from the dust within the transition timescale.
Therefore, the dust must be heated within $\lesssim 10^6$ yr.


\subsection{Transient Heating of Dust}

In the previous section, 
we suggested that the bulk of NH$_3$ gas in the CMZ
must be released from the dust that was previously heated up.
We attribute this heating process
to the followings possibilities: 
interstellar shocks, direct dust collisions, a starburst, and ambipolar diffution.
Considering these mechanisms,
we suggest that
the interstellar shocks caused by a supernova and 
the direct dust collisions would be possible mechanisms
for our transient dust heating scenario.

We first consider the passage of an interstellar shock (\cite{ume99}).
For galactic disk objects, 
the enhancement of the ortho-to-para ratio due to an interstellar shock
is supported by the NH$_3$ observations of
the active and massive star-forming regions (Figure \ref{fig:9}).
In such a source, a strong bipolar outflow is produced, and
it creates a strong shock, although the shocked region
is much more compact than the CMZ.
The interstellar shock can heat the interstellar gas directly.
Although the gas can hardly heat the interstellar dust,
accelerated electrons can heat the dust grain.

In the CMZ, there is considerable evidence 
suggesting the presence of an extended shocked region.
The ubiquity of the intense thermal SiO lines 
(\cite{mar97}; \cite{han06}) and
many shell structures on the $l$-$v$ map observed in the CS line (\cite{tsu99})
are attributed to be caused by the strong shocks passing through the entire CMZ.
These shocks can transiently heat the dust.

What is the origin of the shock? 
Multiple supernovae (SN) shocks are the first possibility.
We estimated the timescale at which SN shocks blow through the CMZ
to be $\sim 10^4$ yr using the multi-phase interstellar gas model of \citet{mck77}.
In this estimation,
we used an SN rate of $10^{-11}$--$10^{-9}$ pc$^{-3}$ yr$^{-1}$ 
(\cite{mun04}; \cite{sch06}), and 
the H atom density at the SN remnant interior of 
$n_{\rm h} = 10^{-2}$ cm$^{-3}$ 
(\cite{mck77}).
In the Galactic center region,
the H atom density of the ambient interstellar medium is larger than 
that in the galactic disk region,
the H atom density at the SN remnant interior may also be larger.
However, even if $n_{\rm h} = 10^{2}$ cm$^{-3}$, 
the timescale is $\sim 10^5$ yr.

Galactic bar shock is the second possibility.
The molecular gas in the CMZ should be strongly affected by
a non-axisymmetric potential.
The gas forms a bar-like structure, 
and is undergoing a strong non-circular motion (\cite{saw04}).
In such a case, 
a strong shock is expected at the leading edge of the bar
(e.g., \cite{sor76}; \cite{ath92}).
This shock is 
observed in extra galaxies (e.g. \cite{han90}; \cite{use06}).
We estimated the timescale 
at which the galactic bar shock passes through the entire CMZ
to be $4 \times 10^6$ yr which is half of the rotation period 
at a radius of 250 pc and a velocity of 200 km s$^{-1}$.
However, the absence of systematic structure 
in the ortho-to-para ratio suggests that
the galactic bar does not produce the fresh NH$_3$.

Cloud-cloud collisions are also a possible trigger of the shock wave.
The energy of turbulent motions in the clouds within 300 pc of the center
is $10^{53}$ erg (\cite{gus85}).
This energy is larger than the thermal energy required to heat the dust to 100 K,
$10^{51}$ erg, which is estimated from 
the total dust mass in the central 400 pc of $5.3 \times 10^{7}$ \MO (\cite{pie00}),
and the specific heat of graphite of $1.3 \times 10^8$ erg K$^{-1}$ g$^{-1}$.
Cloud-cloud collisions may heat the dust surface by the 
transfer of energy through the shock wave.
However, the timescale of cloud-cloud collisions is one order of magnitude longer than
the transition timescale of NH$_3$, $\sim 10^6$ yr, as shown below.
The collision timescale is estimated by
$t_{\rm col} = 1 / (n_{\rm c} \sigma_{\rm c} v_{\rm c})$,
where $n_{\rm c}$ is the number density of clouds,
$\sigma_{\rm c}$ is the collisional cross section,
and $v_{c}$ is the relative velocity.
The number density of clouds is estimated from the total molecular mass 
in the entire CMZ of (3--8) $\times 10^7$ \MO (e.g. \cite{tsu99})
and a cloud mass of $2 \times 10^3$ \MO. 
We use a cloud radius of 1 pc and a H$_2$ number density of $10^4$ cm$^{-3}$ 
in this estimation
because clouds with a size of several pc are 
observed in the CS maps of \citet{tsu99}, and
the critical density of CS is $10^{4}$ cm$^{-3}$.
The collisional cross section is $\sigma_{\rm c} \sim 10^{37}$ cm$^2$ 
at a cloud radius of 1 pc.
The relative velocity of clouds is assumed to be $v_{\rm c} \sim 10$ km s$^{-1}$.
The timescale of cloud-cloud collisions is 
estimated to be $\sim 10^7$ yr from these values.
The comparison of the timescales suggests that
cloud-cloud collisions cannot heat the dust frequently enough.

The direct collision of dust particles is the second mechanism.
The dust surface is heated to $\sim 10^{3}$ K 
when all of the collisional energy is transferred 
to the heating energy of dust.
In this mechanism,
the dust particles must collide within
the conversion timescale of ortho-NH$_3$ and para-NH$_3$.
The collision timescale is estimated by
$t_{\rm col} = 1 / (n_{\rm d} \sigma_{\rm d} v_{\rm d})$,
where $n_{\rm d}$ is the number density of dust particles,
$\sigma_{\rm d}$ is the collisional cross section,
and $v_{d}$ is the turbulent velocity.
The total dust mass within the central 400 pc is $5.3 \times 10^{7}$ \MO (\cite{pie00}),
and the mass of a dust particle is estimated to be $6 \times 10^{-12}$ g
for a dust particle radius of 1 $\mu$m and mass density of charcoal powder of 1.5 g cm$^{-3}$.
Therefore, the number density is estimated to be $n_{\rm d} \sim 10^{-10}$ cm$^{-3}$.
This is an average value over the entire CMZ.
The collisional cross section is $\sigma_{\rm d} \sim 10^{-8}$ cm$^2$
at a radius of 1 $\mu$m.
The turbulent velocity is assumed to be $v_{\rm d} \sim 10$ km s$^{-1}$.
Using these values,
we estimated $t_{\rm col} \sim 10^{4}$ yr.
This timescale is upper limit. If the dust is concentrated
in the core of each cloud with the same turbulent motion,
$n_{\rm d}$ is larger and it gives shorter $t_{\rm col}$.
Before the grains collide directly, 
the gas clouds that contain dust particles collide with each other.
In this case, the dust particles can slow down before the direct collision.
However they are also affected by gas drag, which heats up the dust.
Therefore, the conclusion remains valid, unless the dust particles are
tightly bounded in the molecular clouds by the gas drag.
In this case, the discussion should be the same as that of cloud-cloud collisions.
The comparison of timescales suggests that
the direct collision of dust particles is a possible heating mechanism.

Another heating mechanism is a transient starburst.
When a dust particle is heated up to 100 K by strong UV radiation,
NH$_3$ molecules are released in the interstellar space.
There is substantial evidence that indicates
a burst of star formation in the Galactic center.
However, this occurred $10^7$--$10^9$ years ago
(\cite{yas08} and reference therein).
This timescale is much longer than the transition timescale of NH$_3$. 
Therefore, a transient starburst could not have heated the dust. 

The last mechanism is an ambipolar diffusion.
\citet{hut93} suggested that the gas can be heated up to $\sim$ 200 K
by collisions between neutrals and ions
under a magnetic field strength of 500 $\mu$G. 
However, this mechanism does not work well in our transient heating scenario.
The energy of collision between the dust particle and an ion is
fifteen order of magnitude smaller than that of direct dust collision.
The collisional energy is too small to heat the dust 
to a sufficient temperature to release the NH$_3$, 100 K.
Therefore, the dust could not been heated by ambipolar diffusion,

Our model requires the heating to have occurred within $\lesssim 10^6$ yr.
Therefore, SN shocks, galactic shocks, and direct dust collisions are possible.
However, our estimation of the cooling timescale of molecular gas,
which is described in the next subsection, 
indicates that the timescale should be $\sim 10^5$ yr.
Therefore, 
a galactic shock that occurred $\sim 10^6$ yr ago is not preferable.
We therefore conclude that 
SN shocks and direct dust collisions would be possible mechanisms
for our transient dust heating scenario,
and both mechanisms can produce fresh NH$_3$ from the dust grains.

We expect that there should be the difference of the spatial distributions
in the SN shocks and the direct dust collisions.
In the case of the SN shocks, the dust would be heated in the shocked layer.
In the case of the direct dust collisions, individual heating spot 
is the fairly smaller scale of dust particles.
These minor heated dusts are mixed with the major cool dusts in the observed beam,
therefore they are uniformly observed as cold dusts.
Submillimeter wave observations with very high resolution
could find the differences of the distributions between two mechanisms.


\subsection{Cooling Timescales of Gas and Dust}

NH$_3$ is released from the dust grains to the gas-phase
at a sublimation temperature of $\sim 100$ K.
Therefore, the dust should be heated up to
to $\sim 100$ K in the case that
NH$_3$ molecules are released from the dust.
The observed gas temperature is close to this temperature.
However, the observed dust temperature is much lower.

To resolve this discrepancy, 
we hypothesize that the interstellar matter in the 
Galactic center region is not in thermal equilibrium.
The dust is believed to heat the gas through the photoelectric process,
however, the gas cannot heat the dust.
Therefore, our hypothesis would be confirmed if
the cooling timescale of gas is much longer than
that of dust.

We estimated the timescale in which the gas is cooled 
from 100 to 10 K.
The main cooling process of the gas in this temperature range is CO emission.
The cooling timescale of gas is estimated to be $\sim 10^{5}$ yr
from the thermal energy of $Q = 5/2 k \Delta T \sim 10^{-14}$ erg and
the CO cooling rate of $\sim 10^{-27}$ erg s$^{-1}$ (\cite{heg07}).

We estimated the timescale in which the dust is cooled 
from 100 to 10 K.
The major cooling process of the dust in this temperature range is
black body radiation.
The thermal energy of the dust with temperature $T(t)$ at time $t$
is given by
\begin{eqnarray}
Q(t) = \frac{4 \pi}{3} a^3 \rho c T(t) 
\label{eq:1}
\end{eqnarray}
where $a$, $\rho$, and $c$ denote 
the radius, density, and specific heat of dust.
The cooling rate of the dust
by the black body radiation is given by
\begin{eqnarray}
\frac{dQ(t)}{dt} = - 4 \pi a^2 \sigma [T(t)]^4
\label{eq:2}
\end{eqnarray}
where $\sigma$ is the Stefan-Boltzmann constant.
Using these equations, we found
the cooling timescale of dust 
from 100 to 10 K
to be $t \sim 10^{-3}$ yr for
$a = 1 \mu$m, 
$\rho = 1.5$ g cm$^{-3}$ as the mass density of charcoal powder, and
$c = 1.3 \times 10^8$ erg K$^{-1}$ g$^{-1}$ as the specific heat of the graphite.

The cooling timescale of the gas is 7 orders of 
magnitude longer than that of the dust.
Therefore, transient heating of the interstellar matter occurred in the past, 
and the observed temperature difference between the gas and the dust can be
observed soon after the heating.
This may explain why temperature of molecular gas 
is much higher than that of the dust.


\subsection{Individual Clouds}

\subsubsection{Overview of Clouds in the CMZ}

We estimated the physical parameters of the identified 13 clouds
under the optically thin assumption.
The calculated rotational temperatures, column densities, and ortho-to-para
ratios are listed in Table \ref{tab:3}.

With regard to the clouds in the CMZ,
the virial masses observed in the CO line are one order of magnitude larger than 
the mass derived from the CO luminosity 
with a standard conversion factor (\cite{oka98}).
This suggests that the clouds in the CMZ 
may not be gravitationally bound and are instead
in pressure equilibrium with the hot gas or magnetic field in this region.
However, the cloud observed in the NH$_3$ line 
may not be affected by the external pressure
because the NH$_3$ line would trace a different density range.

Therefore, we investigate the virial mass and luminosity mass in the NH$_3$ line.
Using the abundance ratio of $X$(NH$_3$) = $10^{-9}$ (\cite{hut93})
and spherically symmetric geometry,
the H$_2$ number densities are derived from the luminosity masses to be 
$n$(H$_2$) $\sim 10^3$--$10^4$ cm$^{-3}$, as given in Table \ref{tab:3}.
These values are consistent with the NH$_3$ critical density,
suggesting that the estimated luminosity masses are valid.
We estimated the luminosity mass, $M_{\rm lum}$, from the integration of 
the column densities,
and the virial mass, $M_{\rm vir}$, 
from the apparent size with the beam deconvolution and the observed line width.
We summarize these values of the 13 clouds in Table \ref{tab:3}
and show the $M_{\rm lum}$-$M_{\rm vir}$ plot in Figure \ref{fig:10}.
The luminosity mass and virial mass of nine clouds 
are consistent with an order of magnitude.
This implies that most of the molecular clouds traced 
by the NH$_3$ line are roughly in virial equilibrium.
The external pressure affects only the less dense gas traced by the CO line.

Only four clouds (ID 1, 3, 6, and 8)
appear to not be in the virial equilibrium,
although 
there are reasons in each case, as given below.
In the $l$-$v$ diagram, 
these clouds do not lie on the main ridge 
that indicates the main rotating component of the Milky Way Galaxy.
The rotational temperature and the ortho-to-para ratio of these clouds 
are different from those of the other clouds. 
We summarize the features of these four unvirialized clouds as follows.

\subsubsection{Cloud 1}

Cloud 1 is located at the western edge of the observed area 
and at $v_{\rm LSR} \simeq 0$ km s$^{-1}$.
The velocity and angular extent suggest
that cloud 1 may be located in the Galactic disk.
The annihilation source 1E1740.7-2942 
is located close to the cloud in the sky.
\citet{bal91} assumed that the source is not 
associated to the 0 km s$^{-1}$ component
because it should not be in the Galactic center region.
However, the rotational temperature of 
$45 \pm 4$ K is higher than that of the disk clouds
and close to that of the clouds in the Galactic center.
Therefore, we conclude that
cloud 1 may be located at the Galactic center region.
In a higher-resolution map in the CS line (\cite{tsu99}),
two clouds are observed at the position of cloud 1.
In the case that these clouds are gravitationally unbound,
our estimation of the virial mass is overestimated and 
the large discrepancy between the two masses is not real.

\subsubsection{Cloud 3}

Cloud 3 is located at Sgr C in the sky.
The peak velocity in the NH$_3$ line of $v_{\rm LSR} \simeq -53$ km s$^{-1}$ is 
close to 
the central velocity in the H91$\alpha$ and H109$\alpha$ recombination lines of
$v_{\rm LSR} \simeq -59$ km s$^{-1}$ (\cite{pau75}; \cite{ana89}). 
This suggests that the Sgr C H\emissiontype{II} region 
is physically associated.
It interacts with
the molecular cloud and ionizes a part of it.
The rotational temperature of $45 \pm 3$ K is 
higher than that of other clouds in the CMZ.
This suggests that cloud 3 is heated by the H\emissiontype{II} region
and it is expanding rapidly due to the active star formation.

\subsubsection{Cloud 6}

Cloud 6 is located at high Galactic latitude ($b \simeq 0.15$) at 
$v_{\rm LSR} \sim 106$ km s$^{-1}$.
It is redshifted approximately 60 km s$^{-1}$ from the Sgr A 50 km s$^{-1}$ cloud, 
which is located at the same longitude.
This cloud appears as a shell-like structure on the $l$-$v$ map 
in the CS line (\cite{tsu99}).
Soft X-ray (1--3 keV) emission is detected at the position of this cloud (\cite{wan02}).
These suggest that this cloud is associated with a newly identified supernova remnant.
The large virial mass should be due to expansion of this SN remnant.
The kinetic energy is $10^{51}$ erg when cloud 6 
is expanding with a mass of $M_{\rm lum} = 2.1 \times 10^5$ \MO and 
an expansion velocity of $\Delta v/2$ = 17 km s$^{-1}$. 
This energy is comparable to the kinetic energy of the stellar wind from a 
60--100 \MO star (\cite{abb82}) and less than that of a hypernova explosion
(\cite{iwa98}).

\subsubsection{Cloud 8}

Cloud 8 is the same structure reported in \cite{nag07}.
The rotational temperature and the number density of this cloud 
are higher than those of other clouds.
The location of this cloud is close to that of
the \timeform{0.9D} anomaly reported from the observation in CO line (\cite{oka07}).
The features of high temperature and density are similar to the \timeform{0.9D} anomaly.
Although the highly blue-shifted component ($v_{\rm LSR} = -120$ km s$^{-1}$) 
is detected in CO line, we could not detect this component.

This cloud is unresolved by the Kagoshima 6 m telescope beam.
However, it could be resolved by a high resolution observation 
using the Kashima 34 m telescope.
Figures \ref{fig:11} and \ref{fig:12} show the $l$-$v$ and
$b$-$v$ maps obtained using the Kashima 34 m telescope.
In this higher-resolution observation, the cloud 
is resolved into three components.
The peak positions, apparent sizes, and line widths of 
these components are shown in Table \ref{tab:4}.

We call these three components 8a, 8b, and 8c in the order of the galactic longitude.
Cloud 8 has a large virial mass because it is an unbound system.
The rotational temperatures of all components are similar, 
and they are higher than the mean temperature of the CMZ.
The location and velocity of components 8c is close to
the high velocity compact cloud, CO 0.88$-$0.08 (\cite{oka07}).
These would be the same structure.
However, the component 8a and 8b are not detected in CO line (\cite{oka07}).
This would be because the CO line is affected by foreground absorption. 
We summarize the physical parameters of three components
that have been obtained under the optically thin assumption 
in Table \ref{tab:5}.
Their virial masses are the same as their luminosity masses within a factor of $\simeq 2$.


\section{Conclusions}

We have presented a map of the major part of the CMZ 
by simultaneous observations in the NH$_3$ $(J,K)$ = (1,1), (2,2), and (3,3) lines 
using the Kagoshima 6 m telescope. 
Our observations are summarized as follows:

\begin{enumerate}

  \item
We obtained the ortho-to-para ratio of NH$_3$ in the CMZ 
to be 1.5--3.5 using the rotation diagram method.
This ratio is higher than that of galactic disk clouds,
and as same as those of the active and massive star forming regions
and the centers of nearby external galaxies.

  \item
An ortho-to-para ratio of 1.5--3.5 corresponds to a formation
temperature of 11--20 K.
The formation temperature is very close to the temperature of dust.
NH$_3$ in the CMZ has been produced on the cold dust grains
and released into the gas phase by SN shocks or direct dust collisions.

  \item
The discrepancy between warm molecular gas and cold dust can be
explained by transient heating of the interstellar media
in the CMZ on a timescale $10^5$ years.
NH$_3$ molecules were released from the dust on this timescale.

  \item
We present the physical conditions of the 13 clouds identified in the NH$_3$ map.
For a majority of the clouds in the CMZ,
the luminosity mass and the virial mass are consistent in an order of magnitude. 

\end{enumerate}


\bigskip

We thank an anonymous referee for very useful comments.
We also thank Dr. Mark Morris for his invaluable comments 
about the direct dust collisions.
We also thank K. Takeda,
a graduate from Kagoshima University,
for his technical support in the observations.
We acknowledge S. Takano and T. Umemoto, NRO, for their scientific advice.
We also acknowledge E. Kawai, NiCT, and T. Miyaji and K. Miyazawa, NAOJ, 
for their technical supports in the observations.
T.O. was supported by a Grant-in-Aid for Scientific Research 
from the Japan Society for the Promotion of Science (17340055).
T.H. thanks the Japan Society for the Promotion of 
Science for the financial support provided by 
the JSPS Grant-in-Aid for S (17104002).



\clearpage

\begin{table*}[h]
\begin{center}
\caption{Molecular clouds toward the Galactic center identified in the NH$_3$ line.}
\label{tab:1}
 \begin{tabular}{rrrrrrrl} 
  \hline \hline
  \multicolumn{1}{c}{ID}  &     
  \multicolumn{1}{c}{$l$} &                           
  \multicolumn{1}{c}{$b$} &
  \multicolumn{1}{c}{$v_{\rm LSR}$} &
  \multicolumn{1}{c}{$\Delta l$} &   
  \multicolumn{1}{c}{$\Delta b$} &
  \multicolumn{1}{c}{$\Delta v$} &
  Comments \\
  &
  \multicolumn{1}{c}{(\timeform{D})} &                           
  \multicolumn{1}{c}{(\timeform{D})} &                           
  \multicolumn{1}{c}{(km s$^{-1}$)} &
  \multicolumn{1}{c}{(\timeform{D})} &   
  \multicolumn{1}{c}{(\timeform{D})} &
  \multicolumn{1}{c}{(km s$^{-1}$)} &
  \\
\hline
  1 & $-0.86$ & $-0.11$ & $-10$ & ...  & 0.21 & 53 & at the western edge of the observed area \\
  2 & $-0.86$ & $ 0.00$ & $141$ & 0.20 & 0.17 & 31 & a part of EMR \\
  3 & $-0.51$ & $-0.15$ & $-53$ & 0.17 & 0.16 & 39 & associated with Sgr C \\
  4 & $-0.11$ & $-0.07$ & $ 21$ & 0.21 & 0.16 & 49 & Sgr A 20 km s$^{-1}$ cloud \\
  5 & $ 0.07$ & $-0.08$ & $ 48$ & 0.23 & 0.15 & 49 & Sgr A 40 km s$^{-1}$ cloud \\
  6 & $ 0.07$ & $ 0.15$ & $106$ & 0.16 & 0.19 & 33 & \\
  7 & $ 0.74$ & $-0.13$ & $ 39$ & 0.34 & 0.27 & 65 & Sgr B cloud complex \\
  8 & $ 0.78$ & $-0.12$ & $ 22$ & ...  & 0.16 & 56 & \timeform{0D.9} wing feature\footnotemark[$*$] \\
  9 & $ 0.97$ & $-0.10$ & $ 80$ & 0.31 & 0.19 & 37 & \\
 10 & $ 1.24$ & $-0.07$ & $ 83$ & 0.43 & 0.35 & 49 & the \timeform{1D.3} cloud complex \\
 11 & $ 1.53$ & $-0.23$ & $-25$ & 0.23 & 0.17 & 37 & \\
 12 & $ 1.60$ & $-0.07$ & $ 50$ & ...  & 0.24 & 29 & \\
 13 & $ 1.70$ & $-0.37$ & $-35$ & 0.18 & ...  & 27 & \\
\hline
\multicolumn{8}{@{}l@{}} {\hbox to 0pt{\parbox{50mm}{\footnotesize
\footnotemark[$*$]
  defined in \cite{nag07}
\par\noindent
}\hss}}
\end{tabular}  
\end{center}
\end{table*}

\begin{table*}[h]
\begin{center}
\caption{The integrated intensity of 13 clouds.}
\label{tab:2}
 \begin{tabular}{rrr}
  \hline
  \multicolumn{1}{c}{ID} &
  \multicolumn{1}{c}{$\int T_{\rm MB} dv$} &
  \multicolumn{1}{c}{fraction\footnotemark[$*$]} \\
  &
  \multicolumn{1}{c}{(K km s$^{-1}$)} &
  \multicolumn{1}{c}{(\%)} \\
  \hline
1  &   90 &  1.8 \\
2  &   12 &  0.2 \\
3  &   59 &  1.2 \\
4  &  305 &  6.1 \\
5  &  434 &  8.7 \\
6  &   58 &  6.1 \\
7  & 1224 & 24.6 \\
8  &  107 &  2.2 \\
9  &  246 &  5.0 \\
10 &  562 & 11.3 \\
11 &  135 &  2.7 \\
12 &  151 &  3.1 \\
13 &   64 &  1.3 \\
\hline
total & 3447 & 69.4 \\
\hline
\multicolumn{3}{@{}l@{}} {\hbox to 0pt{\parbox{50mm}{\footnotesize
\footnotemark[$*$]
   The fraction of the total integrated intensity of the entire observed area 
   ($\int T_{\rm MB} dv = 4972$ K km s$^{-1}$).
\par\noindent
}\hss}}
\end{tabular}
\end{center}
\end{table*}

\begin{table*}[h]
\begin{center}
\caption{Physical properties of the 13 clouds.}
\label{tab:3}
 \begin{tabular}{rrrrrrrrrrr}
  \hline
  \multicolumn{1}{c}{ID}             &
  \multicolumn{1}{c}{$T_{\rm rot}$}  &
  \multicolumn{1}{c}{$N$(ortho)}     &
  \multicolumn{1}{c}{$N$(para)}      &
  \multicolumn{1}{c}{$N$(total)}     &
  \multicolumn{1}{c}{$R_{\rm o/p}$}  &
  \multicolumn{1}{c}{$T_{\rm form}$} &
  \multicolumn{1}{c}{$d$}            &
  \multicolumn{1}{c}{$n$(H$_2$)}     &
  \multicolumn{1}{c}{$M_{\rm lum}$}  &
  \multicolumn{1}{c}{$M_{\rm vir}$}  \\
                                            &
  \multicolumn{1}{c}{(K)}                   &
  \multicolumn{3}{c}{$10^{14}$ (cm$^{-2}$)} &
                                            &
  \multicolumn{1}{c}{(K)}                   &
  \multicolumn{1}{c}{(pc)}                  &
  \multicolumn{1}{c}{$10^4$ (cm$^{-3}$)}    &
  \multicolumn{1}{c}{$10^6$ (\MO)}          &
  \multicolumn{1}{c}{$10^6$ (\MO)}          \\
\hline
1   & $45\pm3$ & $ 1.8\pm0.2$ & $1.69\pm0.09$ & $ 3.5\pm0.3$ & $1.1\pm0.2$ & $\geq24$ & 21 & 0.55             &  1.2\hspace{5pt} &  7.2           \\
2   & $31\pm2$ & $ 2.9\pm0.6$ & $1.08\pm0.07$ & $ 3.9\pm0.6$ & $2.6\pm0.7$ & $13\pm3$ & 13 & 0.98             &  0.56            &  1.6           \\
3   & $46\pm2$ & $ 2.6\pm0.2$ & $2.52\pm0.09$ & $ 5.2\pm0.3$ & $1.0\pm0.1$ & $\geq32$ &  6 & 2.8\hspace{5pt}  &  0.15            &  1.1           \\
4   & $33\pm1$ & $16.1\pm0.8$ & $5.99\pm0.10$ & $22.1\pm0.9$ & $2.7\pm0.2$ & $12\pm1$ &  9 & 8.2\hspace{5pt}  &  1.4\hspace{5pt} &  2.6           \\
5   & $37\pm1$ & $13.0\pm0.5$ & $6.83\pm0.10$ & $19.8\pm0.6$ & $1.9\pm0.1$ & $16\pm2$ & 15 & 4.2\hspace{5pt}  &  3.8\hspace{5pt} &  4.6           \\
6   & $32\pm2$ & $ 3.0\pm0.5$ & $1.35\pm0.07$ & $ 4.4\pm0.5$ & $2.2\pm0.5$ & $14\pm3$ &  8 & 1.9\hspace{5pt}  &  0.21            &  1.0           \\
7   & $37\pm1$ & $26.6\pm0.5$ & $14.2\pm0.11$ & $40.8\pm0.6$ & $1.9\pm0.1$ & $16\pm1$ & 38 & 3.5\hspace{5pt}  & 49\hspace{13pt}  & 20\hspace{8pt} \\
8   & $43\pm1$ & $10.2\pm0.3$ & $6.86\pm0.10$ & $17.0\pm0.4$ & $1.5\pm0.1$ & $20\pm2$ &  4 & 15\hspace{13pt}  &  0.20            &  1.5           \\
9   & $41\pm1$ & $ 7.3\pm0.3$ & $4.88\pm0.09$ & $12.2\pm0.4$ & $1.5\pm0.1$ & $20\pm2$ & 25 & 1.6\hspace{5pt}  &  6.3\hspace{5pt} &  4.3           \\
10  & $33\pm1$ & $10.6\pm0.6$ & $4.36\pm0.09$ & $15.0\pm0.7$ & $2.4\pm0.2$ & $13\pm1$ & 52 & 0.93             & 34\hspace{13pt}  & 16\hspace{8pt} \\
11  & $39\pm2$ & $ 3.7\pm0.3$ & $2.04\pm0.08$ & $ 5.7\pm0.4$ & $1.8\pm0.2$ & $17\pm2$ & 15 & 1.2\hspace{5pt}  &  1.1\hspace{5pt} &  2.6           \\
12  & $31\pm1$ & $11.4\pm0.8$ & $4.36\pm0.09$ & $15.7\pm0.8$ & $2.6\pm0.2$ & $13\pm1$ & 27 & 1.9\hspace{5pt}  &  9.4\hspace{5pt} &  2.8           \\
13  & $33\pm2$ & $ 5.3\pm0.6$ & $2.04\pm0.09$ & $ 7.3\pm0.7$ & $2.6\pm0.4$ & $13\pm2$ & 13 & 1.9\hspace{5pt}  &  1.0\hspace{5pt} &  1.2           \\
\hline
  \end{tabular}
\end{center}
\end{table*}

\begin{table*}[h]
\begin{center}
\caption{Three components of the \timeform{0D.9} wing feature 
         observed using Kashima 34 m telescope.}
\label{tab:4}
 \begin{tabular}{rrrrrrr} 
  \hline \hline
  \multicolumn{1}{c}{ID}  &     
  \multicolumn{1}{c}{$l$} &                           
  \multicolumn{1}{c}{$b$} &
  \multicolumn{1}{c}{$v_{\rm LSR}$} &
  \multicolumn{1}{c}{$\Delta l$} &   
  \multicolumn{1}{c}{$\Delta b$} &
  \multicolumn{1}{c}{$\Delta v$} \\
  &
  \multicolumn{1}{c}{(\timeform{D})} &                           
  \multicolumn{1}{c}{(\timeform{D})} &                           
  \multicolumn{1}{c}{(km s$^{-1}$)} &
  \multicolumn{1}{c}{(\timeform{D})} &   
  \multicolumn{1}{c}{(\timeform{D})} &
  \multicolumn{1}{c}{(km s$^{-1}$)} \\
\hline
 8a & 0.767 & $-0.030$ & 25 & 0.065 & 0.065 & 28 \\
 8b & 0.775 & $-0.098$ & 22 & 0.077 & 0.090 & 29 \\
 8c & 0.854 & $-0.059$ & 10 & 0.061 & 0.067 & 35 \\
\hline
  \end{tabular}  
\end{center}
\end{table*}

\begin{table*}[ht]
\begin{center}
\caption{Physical properties of the three components of the \timeform{0D.9} wing feature.}
\label{tab:5}
 \begin{tabular}{rrrrrrrrrrr}
  \hline \hline
  \multicolumn{1}{c}{ID}  &
  \multicolumn{1}{c}{$T_{\rm rot}$} &
  \multicolumn{1}{c}{$N$(ortho)} &
  \multicolumn{1}{c}{$N$(para)} &
  \multicolumn{1}{c}{$N$(total)} &
  \multicolumn{1}{c}{$R_{\rm o/p}$} &
  \multicolumn{1}{c}{$T_{\rm form}$} &
  \multicolumn{1}{c}{$d$} &
  \multicolumn{1}{c}{$n$(H$_2$)} &
  \multicolumn{1}{c}{$M_{\rm lum}$} &
  \multicolumn{1}{c}{$M_{\rm vir}$} \\
  &
  \multicolumn{1}{c}{(K)} &
  \multicolumn{3}{c}{$10^{14}$ (cm$^{-2}$)} &
  &
  \multicolumn{1}{c}{(K)} &
  \multicolumn{1}{c}{(pc)} &
  \multicolumn{1}{c}{$10^4$ (cm$^{-3}$)} &
  \multicolumn{1}{c}{$10^6 (\MO)$} &
  \multicolumn{1}{c}{$10^6 (\MO)$} \\
\hline
 8a & $40\pm1$ & $14.7\pm0.6$ & $9.98\pm0.18$ & $24.7\pm0.8$ & $1.9\pm0.1$ & $16\pm1$ &  9 & 9.1            & 1.6 & 0.86      \\
 8b & $40\pm1$ & $15.4\pm0.6$ & $10.7\pm0.18$ & $26.1\pm0.8$ & $1.5\pm0.1$ & $20\pm2$ & 12 & 7.3            & 3.0 & 1.2\hspace{5pt} \\
 8c & $37\pm1$ & $19.6\pm1.1$ & $9.11\pm0.21$ & $28.7\pm1.3$ & $2.2\pm0.2$ & $14\pm1$ &  9 & 11\hspace{8pt} & 1.8 & 1.3\hspace{5pt} \\
\hline
  \end{tabular}
\end{center}
\end{table*}


\clearpage

\begin{figure*}[h]
  \begin{center}
    \FigureFile(160mm,100mm){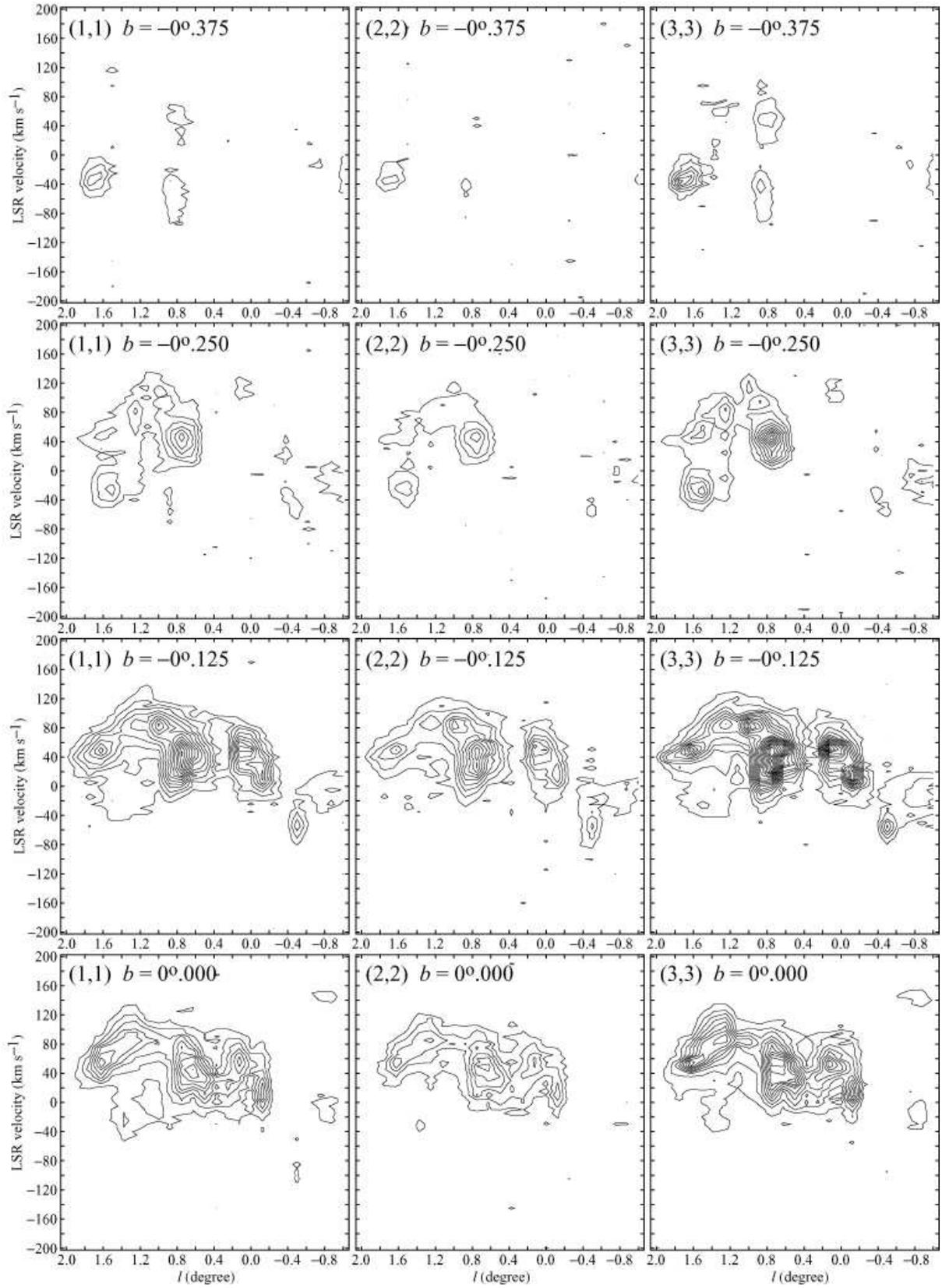}
  \end{center}
  \caption{Longitude-velocity ($l$-$v$) diagrams of
           the NH$_3$ (1,1) (left), (2,2) (middle), and (3,3) (right) lines.
           Both of the lowest contour and the contour interval are 0.1 K.}
  \label{fig:4}
\end{figure*}

\addtocounter{figure}{-1}
\begin{figure*}[h]
  \begin{center}
    \FigureFile(160mm,100mm){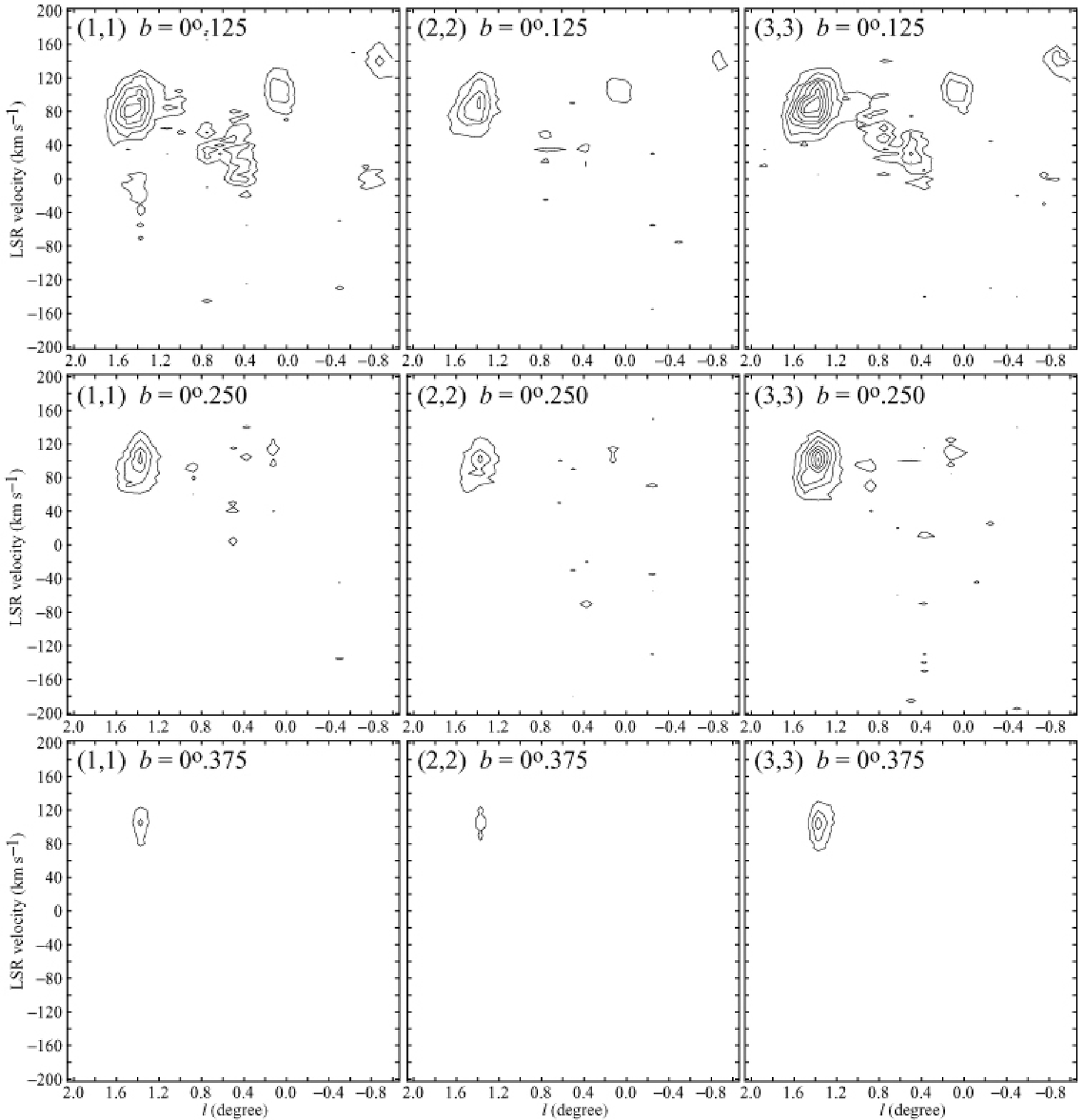}
  \end{center}
  \caption{Continued}
\end{figure*}

\begin{figure*}[h]
  \begin{center}
    \FigureFile(160mm,100mm){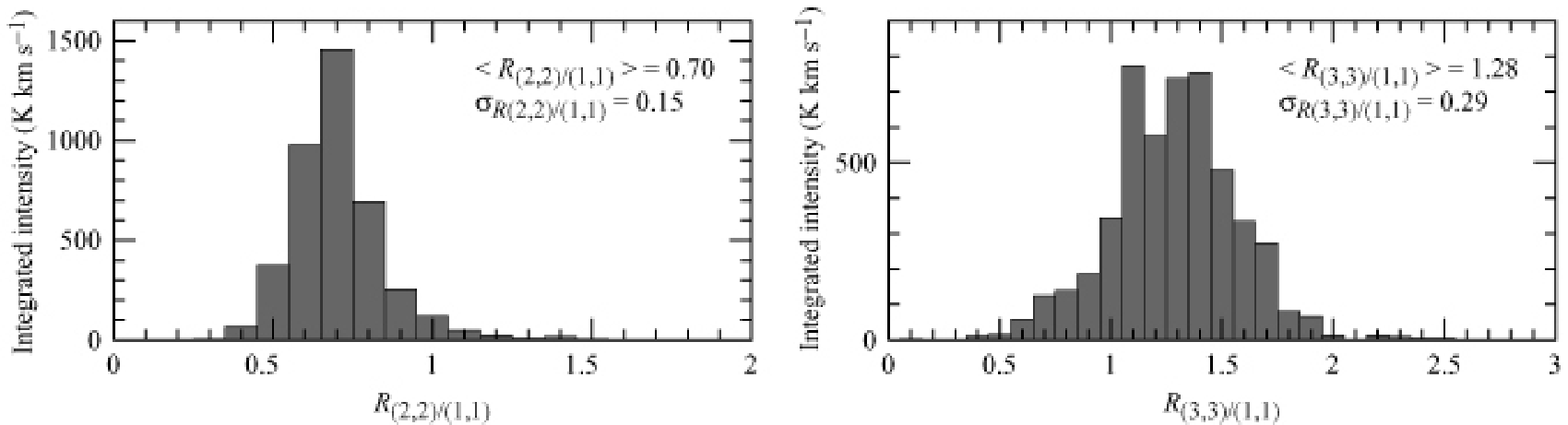}
  \end{center}
  \caption{Histograms of the intensity ratios of (2,2) to (1,1) (left),
           and (3,3) to (1,1) (right).
           We count the $l$-$b$-$v$ pixels at
           which the lines were detected above
           the 3 $\sigma$ level after smoothing at 10 km s$^{-1}$.}
  \label{fig:5}
\end{figure*}

\begin{figure*}[h]
  \begin{center}
    \FigureFile(160mm,80mm){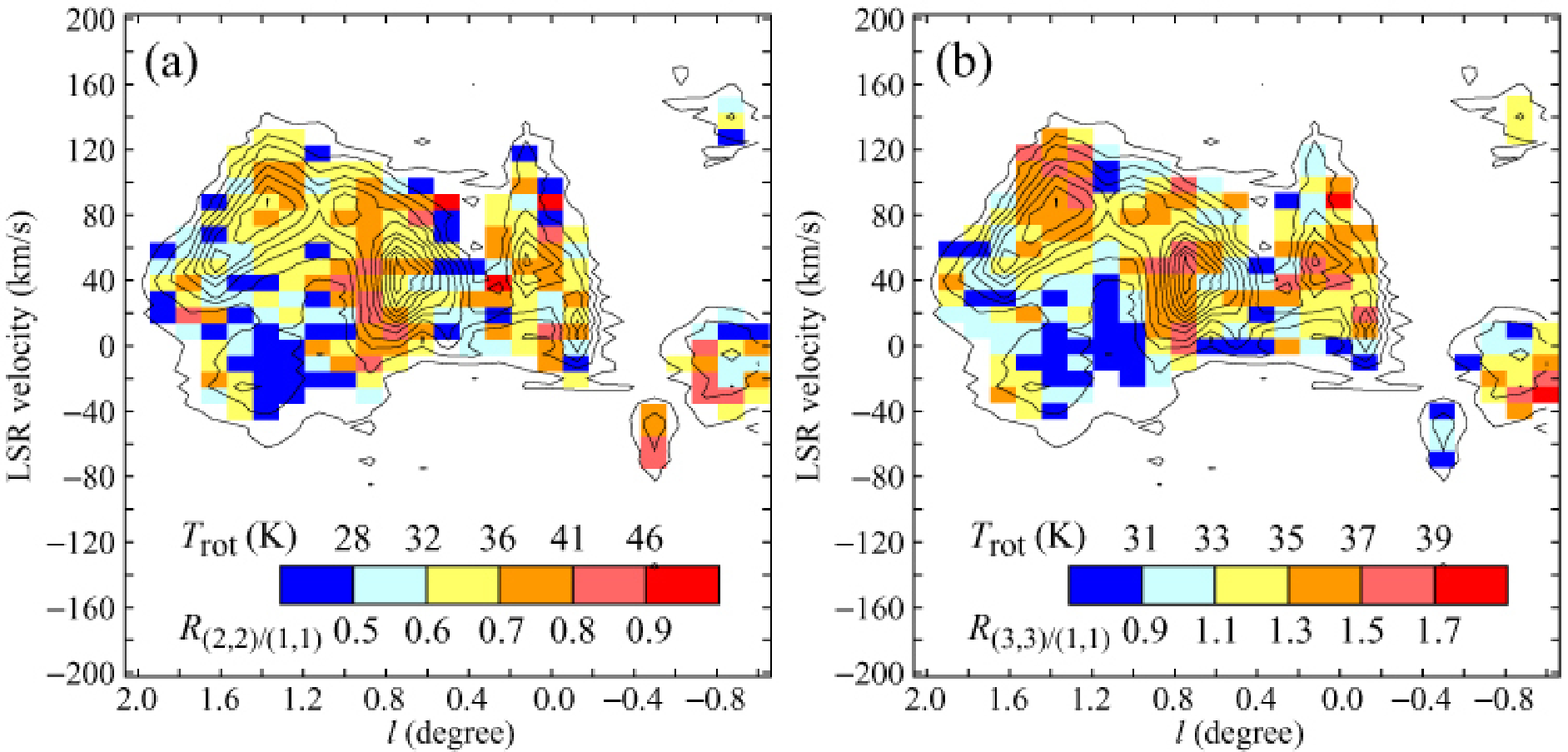}
  \end{center}
  \caption{The distributions of $R_{(2,2)/(1,1)}$ (a)
           and $R_{(3,3)/(1,1)}$ (b) 
           superimposed on the NH$_3$ (1,1) emission integrated over 
           the entire observed latitude.
           $T_{\rm rot}$ in (a) is derived for the optically thin case.
           $T_{\rm rot}$ in (b) is derived for the optically thin case
           and an ortho-to-para ratio of 2.0.}
  \label{fig:6}
\end{figure*}

\begin{figure*}[h]
  \begin{center}
    \FigureFile(80mm,100mm){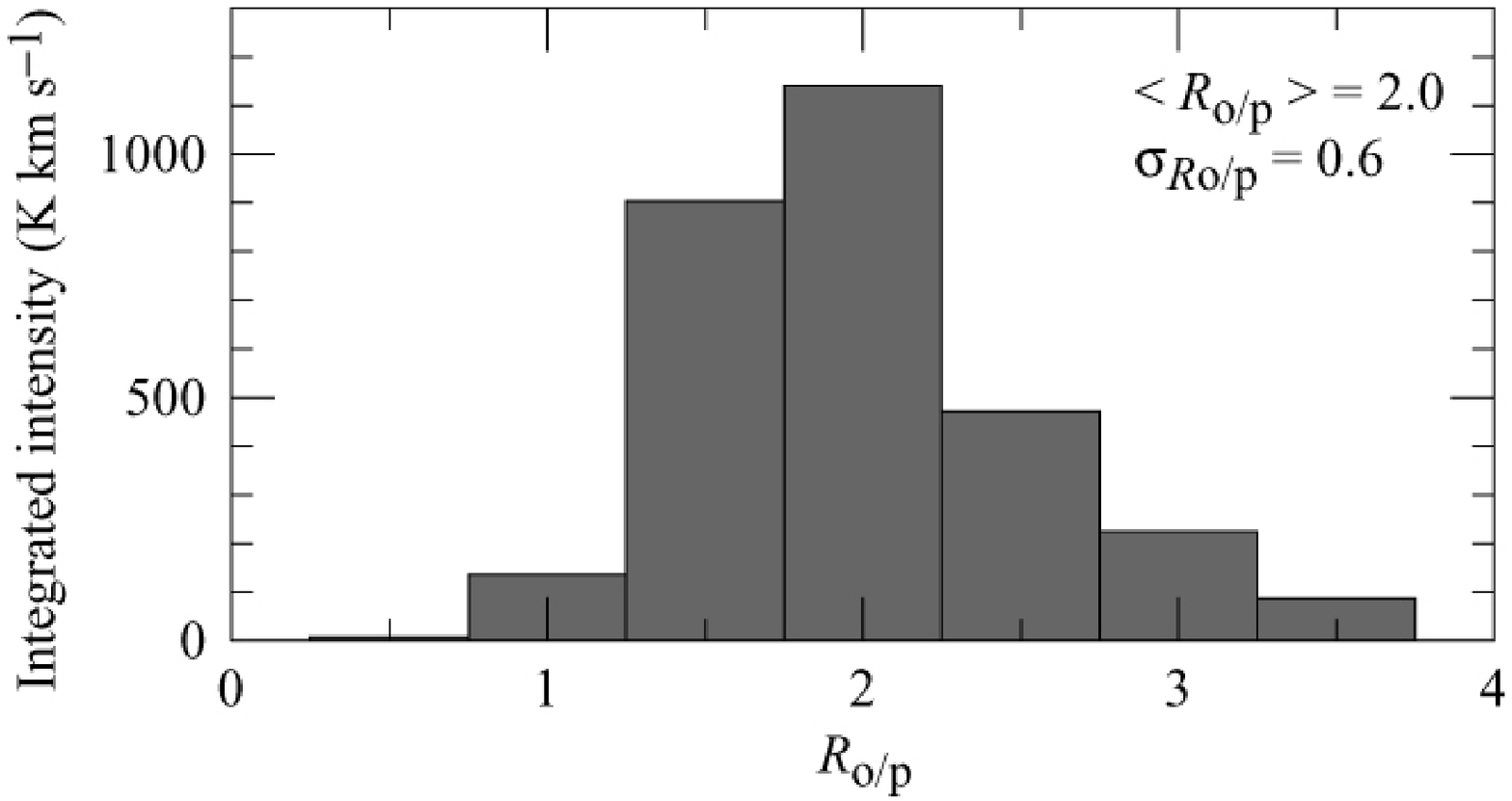}
  \end{center}
  \caption{The histogram of the ortho-to-para ratio.
           We count the $l$-$b$-$v$ pixels at
           which the lines were detected above
           the 10 $\sigma$ level after smoothing at 10 km s$^{-1}$.}
  \label{fig:7}
\end{figure*}

\begin{figure*}[h]
  \begin{center}
    \FigureFile(80mm,100mm){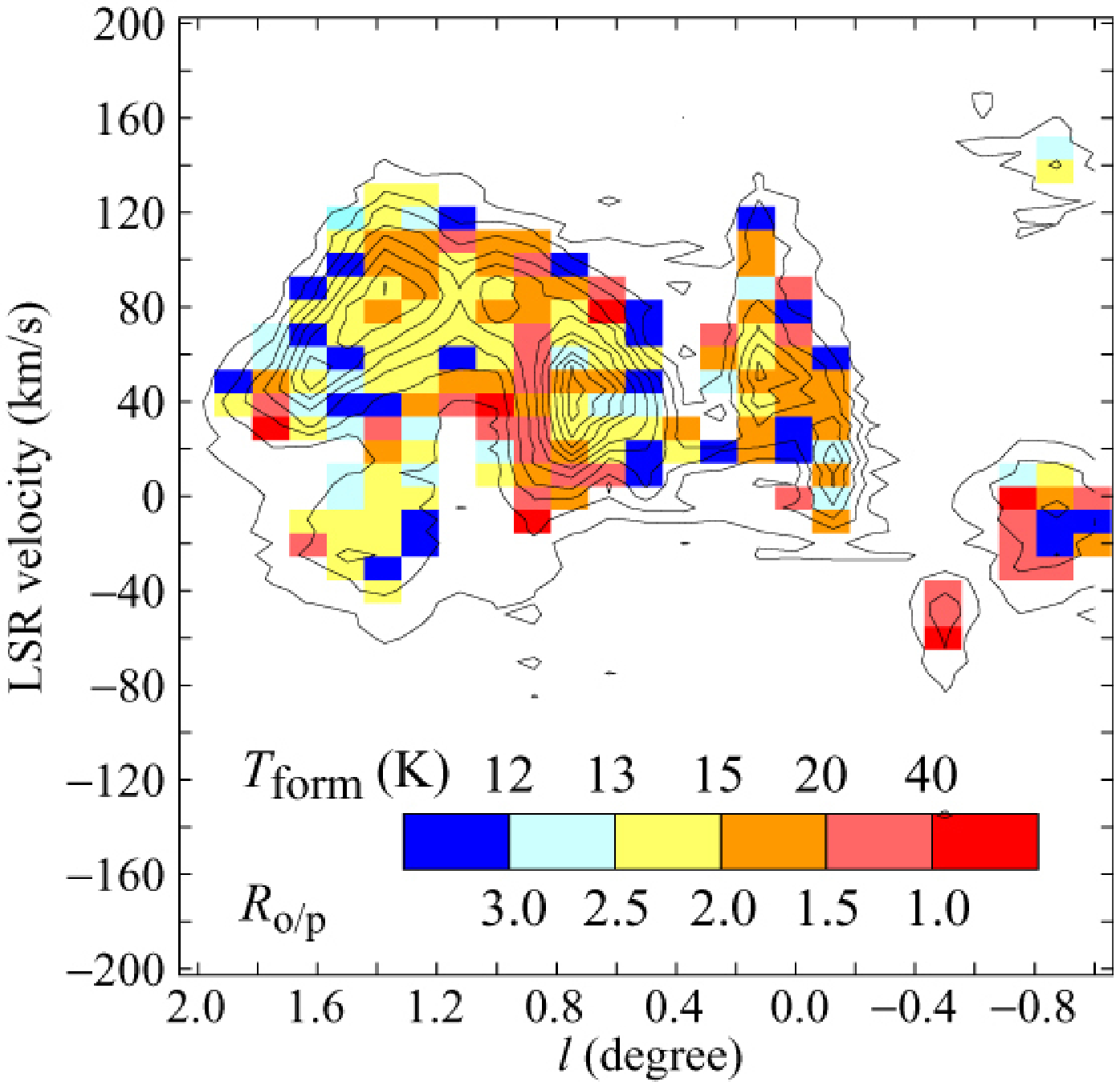}
  \end{center}
  \caption{The distributions of $R_{\rm o/p}$ 
           superimposed on the NH$_3$ (1,1) emission integrated over 
           the entire observed latitude.}
  \label{fig:8}
\end{figure*}

\begin{figure*}[h]
  \begin{center}
    \FigureFile(80mm,100mm){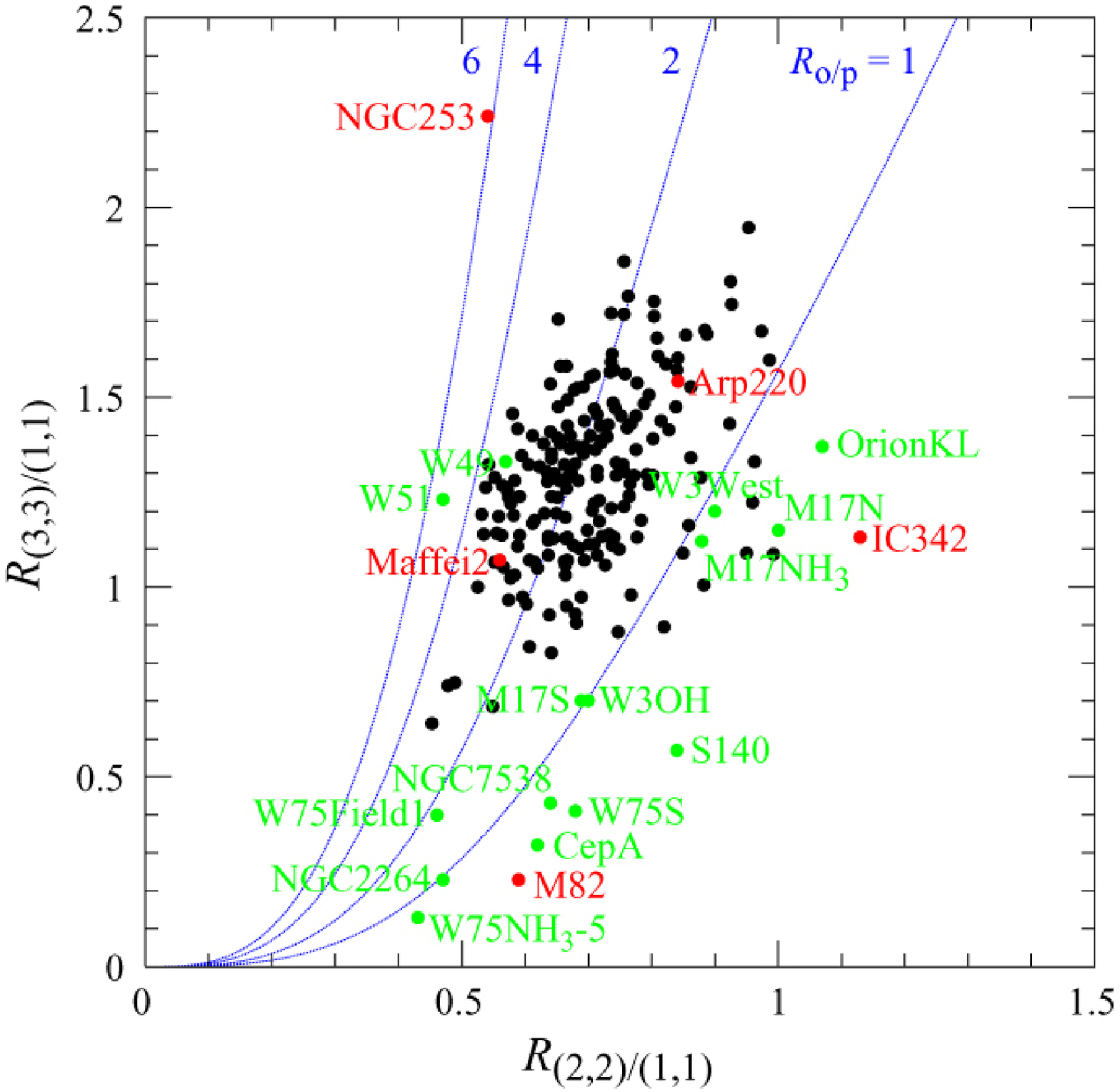}
  \end{center}
  \caption{The $R_{(2,2)/(1,1)}$--$R_{(3,3)/(1,1)}$ correlation plot of
           the CMZ (black), galactic disk clouds (green), and 
           the central region of some external galaxies (red).
           For the CMZ, the spectral data detected over the 10 $\sigma$ level
           after smoothing at 10 km s$^{-1}$ are plotted.
           The data of galactic disk clouds and external galaxies are referred to
           from the following;
           Cep A \citep{ho82}; 
           M17 N,     
           M17 NH$_3$, and
           M17 S \citep{gus88};
           NGC 2264 and 
           NGC 7538 (\cite{nag08} in prep.);
           Orion KL \citep{ho79};
           S140 \citep{mau85};
           W3 West and
           W3 OH \citep{tie98};
           W49 (\cite{nag08} in prep.);
           W51 \citep{mat80};
           W75 Field 1, 
           W75 NH$_3$-5, and
           W75 S \citep{wil90}; 
	   Arp 220 \citep{tak05};
           IC 342 \citep{mau03};
           Maffei 2 \citep{tak00};
           M82 \citep{wei01};
           NGC 253 \citep{tak02};
           Arp 220 is plotted from the ratio of optical depths in the absorption lines.}
  \label{fig:9}
\end{figure*}

\begin{figure*}[h]
  \begin{center}
    \FigureFile(80mm,80mm){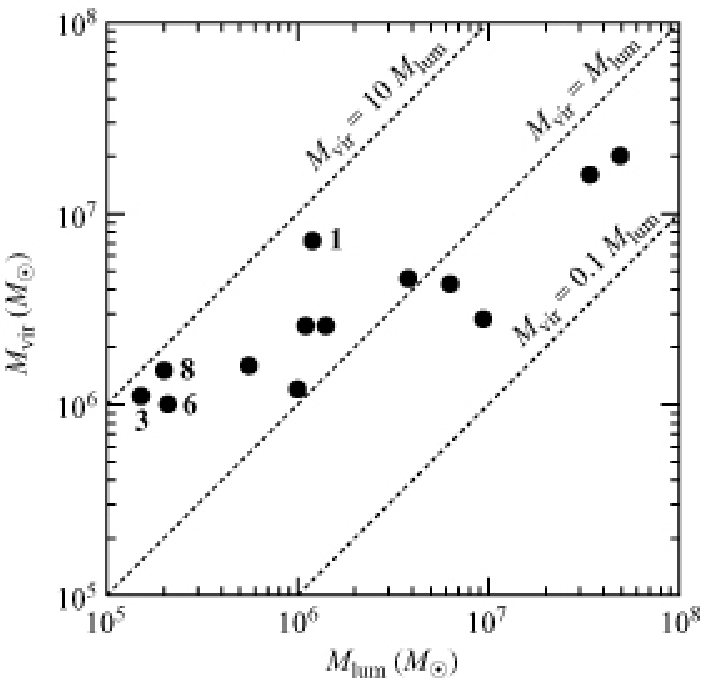}
  \end{center}
  \caption{The luminosity mass-virial mass plot for the 13 clouds.}
  \label{fig:10}
\end{figure*}

\begin{figure*}[h]
  \begin{center}
    \FigureFile(160mm,100mm){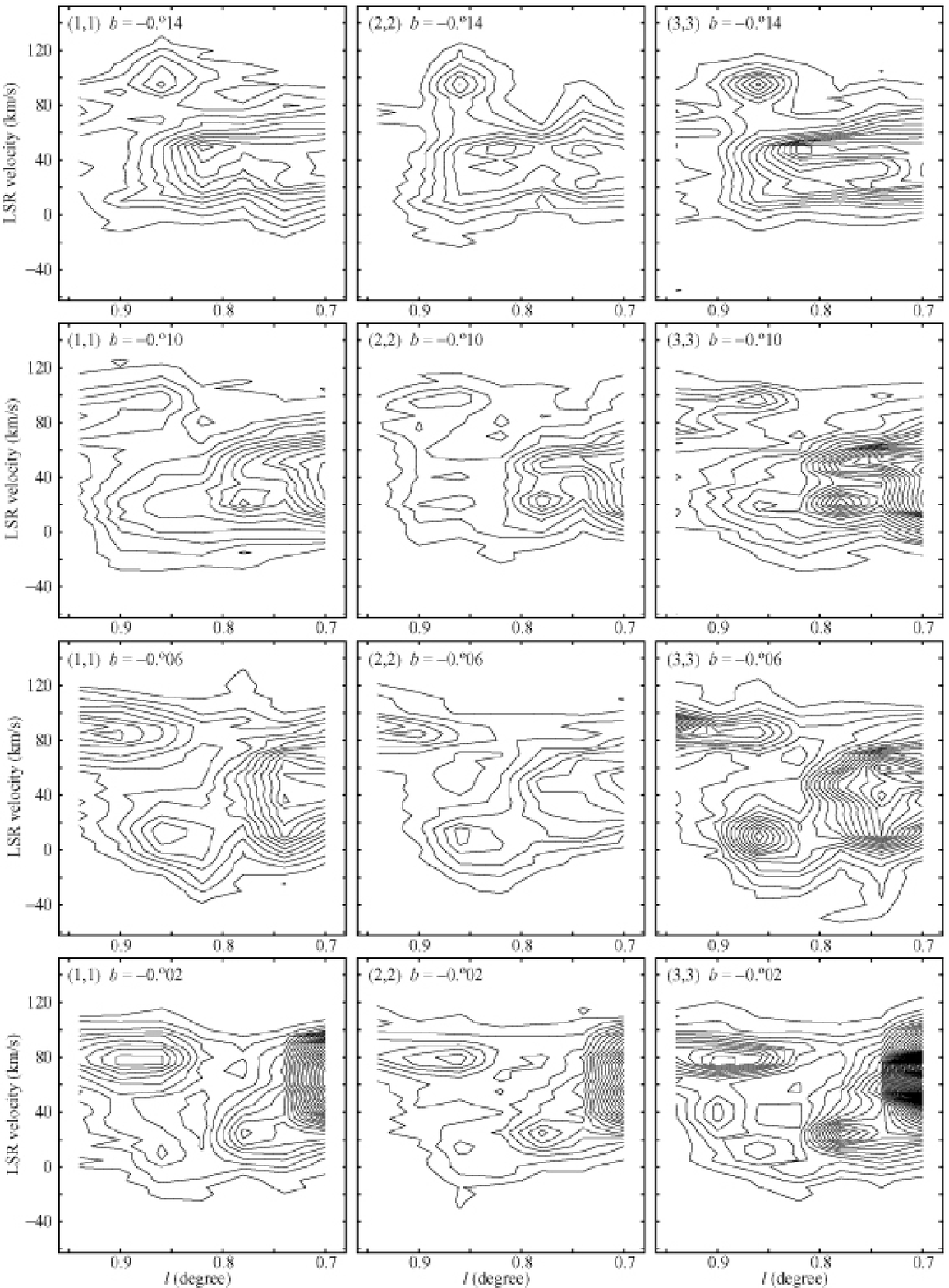}
  \end{center}
  \caption{High-resolution $l$-$v$ diagrams of cloud 8 
           observed using the Kashima 34 m telescope in the
           NH$_3$ (1,1) (left), (2,2) (middle), and (3,3) (right) lines.
           Both the lowest contour and the contour interval are 0.2 K.}
  \label{fig:11}
\end{figure*}

\begin{figure*}[h]
  \begin{center}
    \FigureFile(140mm,100mm){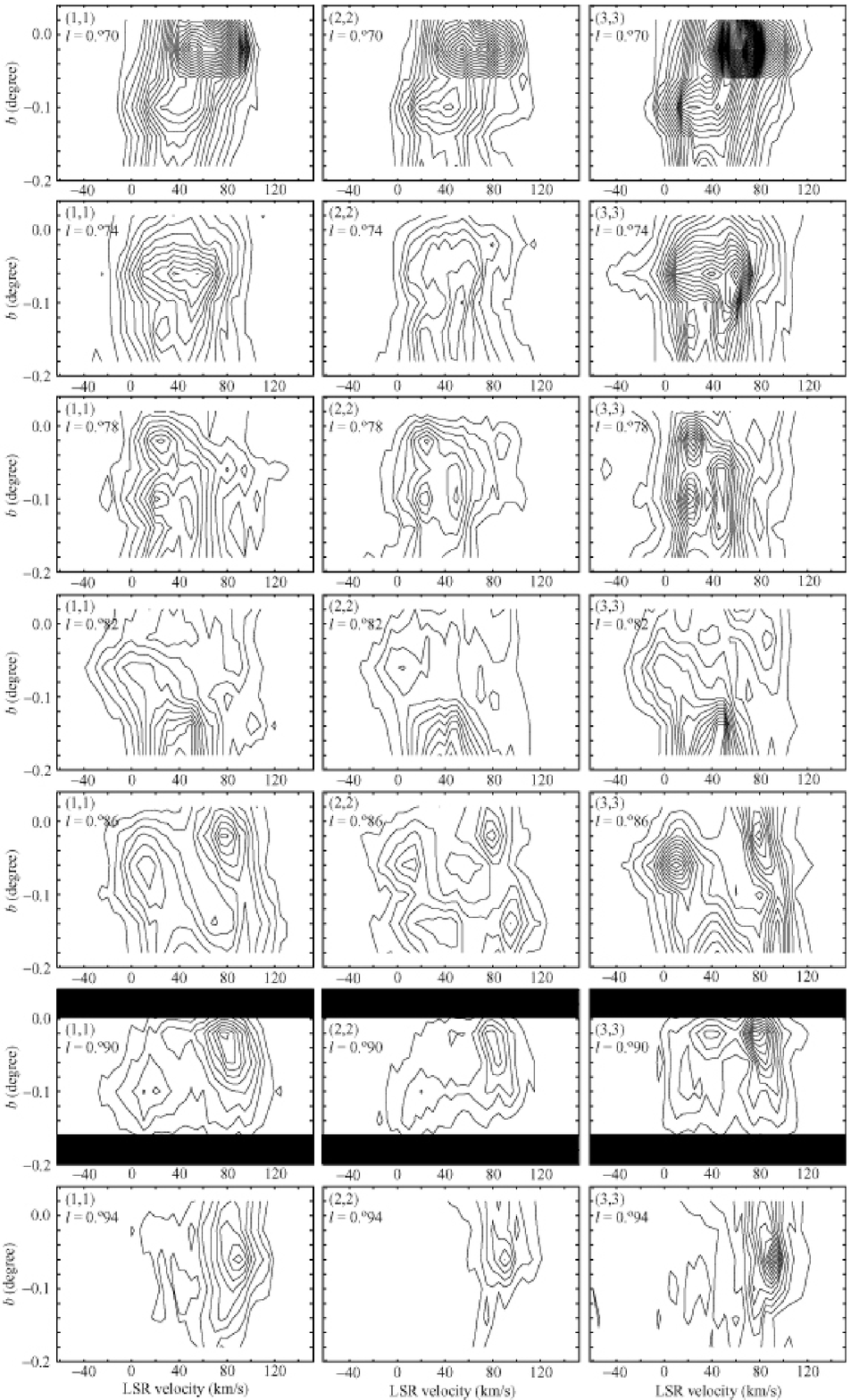}
  \end{center}
  \caption{High-resolution $b$-$v$ diagrams of cloud 8 
           observed using the Kashima 34 m telescope in the
           NH$_3$ (1,1) (left), (2,2) (middle), and (3,3) (right) lines.
           Both the lowest contour and the contour interval are 0.2 K.}
  \label{fig:12}
\end{figure*}

\end{document}